\documentclass[acmsmall,screen,nonacm]{acmart}

\graphicspath{{figures/}}

\usepackage{float}

\AtBeginDocument{%
  }

\usepackage{enumitem}
\usepackage{booktabs}
\usepackage{tabularx}
\usepackage{tikz}
\usetikzlibrary{arrows.meta,positioning}

\begin{document}

\title{RAPID-LLM: Runtime Analysis of Parallel
  Infrastructure for Distributed LLM Training and Inference}

\author{George Karfakis}
\affiliation{%
  \institution{University of California, Los Angeles}
  \city{Los Angeles}
  \state{CA}
  \country{USA}
}

\author{Lime Yao}
\affiliation{%
  \institution{University of California, Los Angeles}
  \city{Los Angeles}
  \state{CA}
  \country{USA}
}

\author{Binglu Chen}
\affiliation{%
  \institution{University of California, Los Angeles}
  \city{Los Angeles}
  \state{CA}
  \country{USA}
}

\author{Faraz Tahmasebi}
\affiliation{%
  \institution{University of California, Irvine}
  \city{Irvine}
  \state{CA}
  \country{USA}
}

\author{Saptarshi Mitra}
\affiliation{%
  \institution{University of California, Irvine}
  \city{Irvine}
  \state{CA}
  \country{USA}
}

\author{Tianyue Pan}
\affiliation{%
  \institution{University of British Columbia}
  \city{Vancouver}
  \state{BC}
  \country{Canada}
}

\author{Hyoukjun Kwon}
\affiliation{%
  \institution{University of California, Irvine}
  \city{Irvine}
  \state{CA}
  \country{USA}
}

\author{Puneet Gupta}
\affiliation{%
  \institution{University of California, Los Angeles}
  \city{Los Angeles}
  \state{CA}
  \country{USA}
}

\renewcommand{\shortauthors}{Karfakis et al.}
\authorsaddresses{}

\begin{abstract}
RAPID-LLM is a unified performance modeling framework for distributed large language model (LLM) training and inference on GPU clusters, without relying on deployment-specific traces or expensive cycle-level simulation for exploration. From a workload and hardware specification, it builds hardware-aware operator-level execution models that capture tiling, memory-hierarchy effects, communication, and memory feasibility under hybrid parallelism. Its backend simulates explicit multidimensional interconnects with congestion-aware routing and support for degraded and failed links, enabling scalable what-if analysis across topology, mapping, and hardware design choices. Across 124 evaluation cases spanning inference and dense, fully sharded, and mixture-of-experts training on A100 and H100 GPUs, RAPID-LLM achieves an overall mean absolute percentage error (MAPE) of 10.0\%. Its network predictions stay within 8\% of ns-3 on representative communication patterns. Case studies demonstrate how RAPID-LLM enables fast, systematic sweeps over hybrid-parallel configurations, quantifies sensitivity to link faults under realistic routing and congestion, and evaluates hypothetical GPU design variants including 3D-stacked HBM-on-GPU scenarios.
\end{abstract}




\maketitle

\section{Introduction}

Today’s large language models (LLMs) are trained and served on clusters of GPU-based systems comprising hundreds to thousands of nodes. Training runs are long and communication-heavy, while inference workloads push for high throughput and tight latency. In both phases, end-to-end performance emerges from coupled interactions between GPU microarchitecture, the memory hierarchy, and the surrounding interconnect fabric. As deployments grow in scale and complexity, it becomes increasingly difficult to predict how changes in networking, workload structure and hardware design will affect training time, inference latency, and overall resource efficiency.

To explore future systems, practitioners increasingly turn to modeling tools that predict end-to-end LLM training and inference behavior using configurable abstractions. Symbolic or analytic workload descriptions coupled to network simulators (e.g., Astra-Sim and STAGE/MLSynth \cite{astrasim,man2025stage,sefiane2025mlsynth}) can represent large distributed jobs and a variety of topologies, but they often reduce compute and memory models to coarse layer-level models. Trace-driven tools (e.g., SimAI and ReaLLM \cite{wang2025simai,peng2025reallm}) ingest measurements from real runs and achieve high realism for the profiled hardware and network, but they are tied to that deployment. Detailed GPU or multi-GPU simulators (e.g., Accel-Sim and GPGPU-Sim \cite{khairy2020accelsim,bakhoda2009gpgpusim}) can model fine-grained behavior, but their high runtime confines them to smaller systems and fewer design points rather than cluster-scale, forward-looking exploration.

To address these limitations, a useful modeling framework must bridge hardware detail and cluster-scale scope. It should capture operator-level compute, cache and memory interactions, and realistic interconnect dynamics, while scaling to large systems and running fast enough to explore many system configurations. It should support both training and inference on a shared hardware description across common LLM parallelization strategies and explicitly model link faults and routing-dependent congestion. To enable future-facing design space exploration, it should accept abstract specifications of GPUs, memory stacks, packaging, and network topologies instead of traces from a specific deployment.

RAPID-LLM addresses these requirements with a frontend that generates hardware-informed, operator-level execution traces from abstract LLM configurations and a backend that executes those traces on fault- and congestion-aware multidimensional networks.

Two examples illustrate why these properties matter. Consider a large LLM training job that uses pipeline parallelism across many GPUs. A single degraded or failed inter-node link can have very different impact on end-to-end throughput depending on its position: in one layout, it lies on a critical cut and stalls the pipeline. In another, most traffic bypasses it and throughput changes little. Predicting these outcomes requires modeling topology, time-varying congestion, and rerouting due to link failures, not just average per-link bandwidth.

As a second example, consider the attention operator in a transformer block. FlashAttention~\cite{flashattn} performs similar arithmetic to a naive implementation but exercises the memory hierarchy very differently: it keeps working set tiles resident in fast on-chip memory, whereas naive attention streams full intermediate tensors from external memory. On real GPUs, these data-movement differences often dominate runtime~\cite{flashattn}. Any tool that only counts FLOPs or uses layer-level bandwidth models will mispredict their relative performance and sensitivity to cache sizes and memory bandwidth. Accurate modeling requires hardware-aware traces that capture multi-level tiling and memory traffic.

This paper makes the following contributions:
\begin{enumerate}
  \item We develop a hardware-aware operator model, centered on an extended-roofline GEMM model that captures GPU tiling, wave scheduling and memory-hierarchy traffic, and use it to produce per-operator Chakra execution traces for LLM training and inference under hybrid parallelism strategies.
  \item We model multidimensional networks with per-link congestion and support for degraded and failed links, enabling large-scale simulation across diverse topologies and fault patterns.
  \item We use these components to build RAPID-LLM, a highly scalable and unified modeling framework, validate it against real deployment measurements with comparisons to representative existing tools, and conduct case studies analyzing packaging, memory configurations, and the impact of link faults and mapping decisions.
\end{enumerate}

The remainder of this paper is organized as follows. Section 2 provides background on LLM workloads, parallelism strategies, cluster interconnects, and link faults at scale. Section 3 reviews existing approaches and motivates the need for a more robust unified modeling framework. Section 4 describes the design of RAPID-LLM, including its constituent components and their extensions. Section 5 presents our validation methodology and comparative evaluation. Section 6 presents case studies that visit hardware and network what-if scenarios that are uniquely enabled by RAPID-LLM. Section 7 concludes with a discussion of limitations and directions for future work.

RAPID-LLM is available as open source at \url{https://github.com/nanocad-lab/Rapid-LLM}.

\section{Background}

\paragraph{Modern LLM deployment on server-class clusters} 
As LLMs scale to trillions of parameters, training and inference increasingly rely on datacenter-scale GPU clusters. For example, Llama3.1-405B was trained on 16{,}384 H100 GPUs with a reported Model FLOPs Utilization (MFU) of only $\approx 38\%$ \cite{dubey2024llama}. At this scale, throughput is often limited less by raw compute than by memory capacity and communication: parameters, activations, optimizer state, and inference-time KV caches dominate device memory usage, stressing HBM capacity and increasing communication pressure \cite{rajbhandari2020zero,kwon2023pagedattention,aminabadi2022deepspeedinference}. How this state is partitioned and communicated directly shapes training throughput and inference latency.

These trends motivate hybrid parallelism strategies and the need for models that capture the interaction between compute, memory, and communication on large GPU clusters.

\paragraph{Parallelism strategies in distributed LLMs}
We focus on six spatial parallelism axes used in large-scale LLM systems:  
\begin{itemize}[leftmargin=*]
  \item \textbf{Data parallelism (DP)} replicates the model across workers and shards the input batch across replicas.
  \item \textbf{Tensor/model parallelism (TP)} partitions individual layers across devices so that each GPU holds only a shard of the weights and activations \cite{shoeybi2019megatron}.
  \item \textbf{Pipeline parallelism (PP)} partitions the layer stack into stages mapped to different GPUs, with microbatches flowing across stages \cite{huang2019gpipe}.
  \item \textbf{Sequence parallelism (SP)}, typically combined with Tensor Parallelism, shards intermediate activations along the sequence dimension and synchronizes around sequence-wide operations \cite{korthikanti2023reducing}.
  \item \textbf{Context parallelism (CP)} assigns different token ranges to different GPU groups and exchanges attention KV tensors between them before attention \cite{liu2023ring}.
  \item \textbf{Expert parallelism (EP)} shards Mixture-of-Experts (MoE) layers by assigning different experts to different GPUs and exchanging token activations via all-to-all routing between experts \cite{lepikhin2020gshard}.
\end{itemize}

In practice, production deployments use hybrid parallelism. For example, the final stage of Llama3.1-405B pre-training runs on 16{,}384 GPUs with TP=8, CP=16, PP=16, and DP=8 to support 131k-token sequences \cite{dubey2024llama}, creating overlapping collective and point-to-point communication. Training consists of a forward pass (FP) and a backward pass (BP) over the network, while inference (prefill and decode) largely reuses FP communication patterns. Table~\ref{tab:parallelism} summarizes the dominant communication patterns introduced by each parallelism axis in Megatron-style implementations \cite{shoeybi2019megatron}.

\begin{table}[t]
\centering
\caption{Dominant communication patterns for common parallelism types in large-scale LLM training. Backward pass communication includes activation and weight gradients. Patterns follow common Megatron-style~\cite{shoeybi2019megatron} configurations and are indicative only. Abbreviations: AG = AllGather, AR = AllReduce, RS = ReduceScatter, A2A = all-to-all, P2P = point-to-point send/recv, blk = structured block exchange.}
\label{tab:parallelism}
\begin{tabular}{@{}lcc@{}}
\toprule
\textbf{Parallelism} & \textbf{Forward Pass} & \textbf{Backward Pass} \\
\midrule
Data (DP) 
    & -- 
    & AR \\

Tensor/model (TP) 
    & AG / AR 
    & AR \\

Pipeline (PP) 
    & P2P 
    & P2P \\

Sequence (SP) 
    & AG / RS 
    & AG / RS \\

Context (CP) 
    & A2A / blk 
    & A2A / blk \\
Expert (EP)
    & A2A
    & A2A \\
\bottomrule
\end{tabular}
\end{table}

\paragraph{Cluster interconnects and topologies}

Modern GPU clusters expose a multi-level communication hierarchy: GPUs within a node are connected by high-bandwidth, low-latency fabrics such as NVLink and NVSwitch, while inter-node communication relies on InfiniBand or RDMA-capable Ethernet \cite{nvidia_nvlink,li2020gpuinterconnect,nvidia_infiniband,nvidia_ethernet}. Logically, the cluster forms a multidimensional network spanning intra-node, intra-rack/pod, and inter-rack/pod dimensions.

At the network level, links are organized into switch-centric Fat-Tree/Clos or direct-connect Mesh and Torus topologies \cite{al2008scalable,dally2004principles}. Collective operations generate heavy traffic on these fabrics, creating congestion that depends on the mapping of parallelism groups to the physical topology. Realistic performance models must therefore capture the hierarchical structure of links, multidimensional contention, and routing-dependent effective bandwidth.

\paragraph{Faults and degraded links at scale} 

Hardware and network instability are a constant reality at LLM training scale \cite{gill2011dcnetfailures}. For instance, Meta reports over 400 unexpected job interruptions during Llama3.1-405B training on a 16k-GPU cluster \cite{dubey2024llama}, with nearly half attributed to hardware or infrastructure failures. More subtle link-level problems include persistent bandwidth degradation or elevated error rates that cause retransmissions \cite{gunawi2018failslow}, which reduce the effective capacity of specific routes \cite{huang2017grayfailure}.

Hybrid parallelism creates tightly coupled communication patterns across many GPUs, so such link-level anomalies on a critical cut can materially reduce effective system performance. Models that assume fully healthy links and static per-link bandwidth cannot capture these effects. Accurately reasoning about performance therefore requires congestion-aware network models that can represent both contention and explicit link faults.

\section{Existing Approaches and Related Work}
Prior work on modeling and simulation for large-scale LLM training and inference spans a wide spectrum of fidelity and scalability. Most systems can be understood as composing a compute frontend that produces a time-ordered stream (or dependency graph) of compute and communication events, and a network/backend model that predicts the cost of those communication events under a chosen topology and contention model, sometimes with a similarly abstract compute backend. We adopt this decomposition to separate what existing tools can represent about LLM workloads from what they can realistically predict about cluster-scale execution.

\paragraph{Compute frontend models}
Compute frontends for LLM training and inference broadly fall into three categories: (1) trace-driven profiling and replay, (2) cycle-level (or cycle-approximate) accelerator simulators, and (3) analytical node-level models.

Trace-driven tools extrapolate execution profiles from real runs, offering high realism but limited ability to explore future hardware and network designs due to their reliance on system-specific measurements. Examples include SimAI \cite{wang2025simai} and Chakra- or ReaLLM-style trace replay \cite{sridharan2023chakra,peng2025reallm}.

Cycle-level simulators provide detailed modeling of accelerator internals, such as GPGPU-Sim and Accel-Sim for GPUs \cite{bakhoda2009gpgpusim,khairy2020accelsim} and ONNXim for NPUs \cite{ham2024onnxim}, but their high runtime limits scalability to cluster-scale LLM studies.

Analytical node-level models use FLOP- and tensor-size-based roofline estimates. MLSynth \cite{sefiane2025mlsynth} and STAGE \cite{man2025stage} follow this approach. However, coarse global scaling reduces accuracy when performance depends on operator shape or memory hierarchy effects. Our frontend remains analytical but incorporates operator-level, hardware-aware memory modeling to bridge roofline models and trace-driven realism.

\paragraph{Network backend models}

Given a stream of communication events from a compute frontend, modeling tools typically rely on one of three network backends: simple analytic models, packet-level simulators, or analytical topology- and congestion-aware models. Astra-Sim \cite{astrasim} exposes all three as swappable backends, a pattern also seen in end-to-end tools.

Simple analytic backends estimate communication time from message size, bandwidth, and latency, and are fast but unable to capture congestion or link-level effects (e.g., ReaLLM \cite{peng2025reallm}, Astra-Sim’s congestion-unaware backend).

At the other extreme, packet-level simulators such as ns-3 \cite{ns3} provide high-fidelity modeling of arbitrary topologies and congestion, but are prohibitively slow for cluster-scale LLM studies.

Between these extremes, analytical congestion-aware backends explicitly model links and contention. Astra-Sim's default congestion-aware analytical backend, however, supports only a single logical dimension and assumes homogeneous, healthy links.

\paragraph{Gaps in existing tools and RAPID-LLM's approach}

Existing modeling tools fall short in three areas. First, compute frontends either rely on deployment-specific traces (limiting hardware exploration) or use coarse roofline models that miss operator-shape and memory-hierarchy effects. Second, network backends either ignore congestion entirely, model only a single dimension, or require prohibitively slow packet-level simulation. Third, no existing tool models link faults (soft degradation or hard failures) and their interaction with routing and congestion.

\begin{figure}[!htbp]

  \centering
  \includegraphics[width=0.725\linewidth]{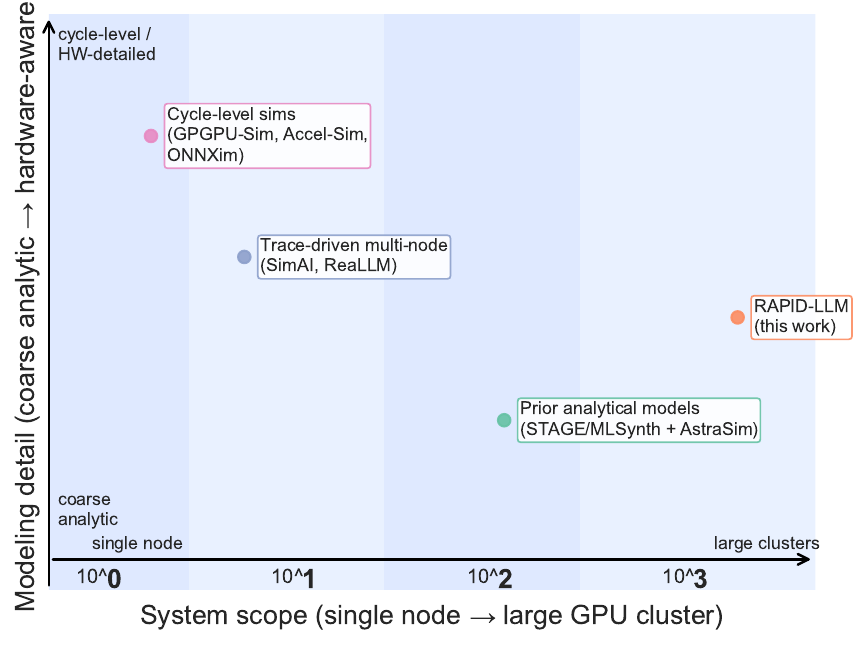}
  \caption{Qualitative positioning of LLM modeling tools along modeling detail and system scope. X axis refers to the number of GPUs in a system. RAPID-LLM targets a regime of operator-level, hardware-aware modeling at cluster scale.}
  \Description{Cartoon scatter plot placing LLM modeling tools on axes of modeling detail versus system scope, with RAPID-LLM in the upper-right region combining operator-level detail with cluster scale.}
  \label{fig:detail-scope-cartoon}
\end{figure}

RAPID-LLM fills this gap with an operator-level, hardware-aware frontend that requires no real-system profiling and a backend that models explicit multidimensional links, per-link congestion, and fault-aware routing at analytical cost, even for thousand-GPU clusters. Figure~\ref{fig:detail-scope-cartoon} positions RAPID-LLM against existing tools, and Table~\ref{tab:tool_comp_table} provides a detailed comparison.

\begin{table}[t]
\centering
\scriptsize
\caption{High-level capability comparison. ``AS'' refers to Astra-Sim \cite{astrasim}. ``Adv. LLM features'' refers to ZeRO-1/2/3 DP, Mixture of Experts (MoE), grouped-query and multi-head latent attention (GQA/MLA), activation recomputation/checkpointing and FlashAttention \cite{flashattn}. ``RL`` refers to roofline-style models.}
\label{tab:tool_comp_table}
\setlength{\tabcolsep}{2.5pt}
\renewcommand{\arraystretch}{1.15}
\begin{tabular}{>{\raggedright\arraybackslash}p{1.9cm}ccccccc}
\hline
\textbf{Comparison Point} &
\textbf{RAPID-LLM} &
\shortstack{\textbf{llm-analysis}\\\cite{li2023llmanalysis}} &
\shortstack{\textbf{STAGE}\\\textbf{+AS}~\cite{man2025stage,astrasim}} &
\shortstack{\textbf{MLSynth}\\\textbf{+AS}~\cite{sefiane2025mlsynth,astrasim}} &
\shortstack{\textbf{LLM-}\\\textbf{Compass}~\cite{zhang2024llmcompass}} &
\shortstack{\textbf{GenZ-LLM}\\\textbf{-Analyzer} \cite{genz}}&
\textbf{Vidur} \cite{vidur}\\
\hline
Training/Inference & \textbf{Both} & Both & Training & Training & Inference & Inference & Inference \\
Adv. LLM features & \textbf{Yes} & Yes & Yes & No & Yes & Yes& Yes\\
HW Compute model & \textbf{Adv. (GPU-aware RL)} & Basic RL & Basic RL & Basic RL & Adv. & Basic RL & Fitted ML model\\
Mem hier. model & \textbf{HBM/L2/SRAM} & None & None & None & Basic (tile) & None & None \\
Network model & \begin{tabular}[t]{@{}c@{}}\textbf{AS + multi-}\\\textbf{dim. model}\end{tabular} & Analytical & Standard AS & Standard AS & Analytical & Standard AS & Fitted ML model\\
Congestion model & \textbf{Multi-dim} & None & Single-dim & Single-dim & None & None& None \\
Multi-dim topologies & \textbf{Yes} & No & Limited & Limited & No & No& No \\
Fault modeling & \textbf{Yes} & No & No & No & No & No & No \\
\hline
\end{tabular}
\end{table}
The next section describes its architecture and components.

\section{RAPID-LLM} 

\begin{figure}[!htbp]
  \centering
  \includegraphics[width=\linewidth]{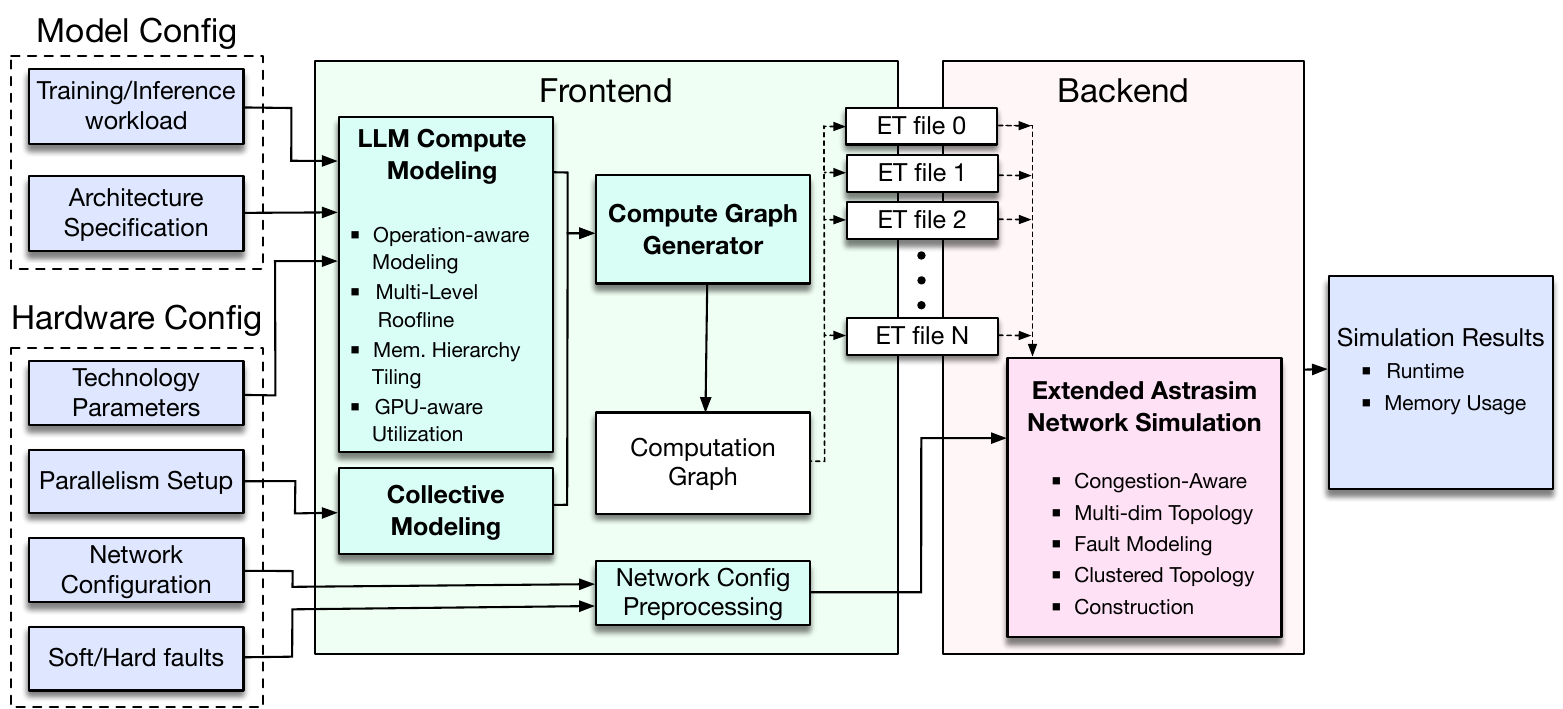}
  \caption{Overview of RAPID-LLM}
  \Description{Block diagram of the RAPID-LLM workflow: LLM workload and hardware specifications feed a frontend that builds an operator-level computation graph and emits per-GPU Chakra traces, which a fault- and congestion-aware network backend executes to produce runtime estimates and breakdowns.}
  \label{fig:rapid-llm-overview}
\end{figure}

Figure~\ref{fig:rapid-llm-overview} shows the overall RAPID-LLM workflow. Each simulation run takes two inputs: an LLM workload specification (model architecture, training or inference request, and parallelism configuration) and a hardware specification (per-device parameters, cluster topology, and optional link faults). The frontend consumes these inputs to build an operator-level computation graph with explicit compute and communication tasks, then emits per-GPU Chakra execution traces. The backend executes these traces on an extended Astra-Sim network model with multidimensional congestion and fault awareness, producing end-to-end runtime estimates and compute/communication breakdowns for both training and inference.

\subsection{Frontend: workload specification and trace generation}
The RAPID-LLM frontend transforms high-level LLM and hardware specifications into an operator-level computation graph and per-GPU Chakra execution traces that drive the network backend.

\paragraph{Workload and hardware specification.}
RAPID-LLM is organized around the same overall framework as DeepFlow~\cite{pal2023deepflow}, with new modeling support for modern decoder-only LLM training and inference. Models are described by parameters such as number of layers, hidden dimension, number of attention heads, number/size of experts, and sequence lengths, in addition to training- or inference-specific hyperparameters such as prefill/decode length and batch size. The deployment is described by a hardware specification (per-GPU compute and memory parameters) and a hybrid parallelism configuration that sets the degrees of DP, PP, CP, TP, and EP and maps them onto the available GPUs. Given these inputs, the frontend synthesizes a per-GPU workload that assigns each operator to a device according to the chosen parallelism strategy.

\paragraph{Computation graph and Chakra traces.}
From this per-GPU workload, RAPID-LLM constructs a central directed acyclic graph (DAG) whose nodes correspond to compute and communication tasks and whose edges encode data and control dependencies. The graph reflects the chosen hybrid parallelism configuration, with tensor, expert and context parallelism sharding intra-layer operators across GPUs and pipeline and data parallelism introducing cross-layer and cross-microbatch dependencies.
For expert-parallel workloads, token routing across experts can be balanced or imbalanced: RAPID-LLM can assign one expert a specified multiple of the balanced-average token load and distribute the residual uniformly across the remaining experts, reflecting the fact that expert utilization is not always balanced in practice~\cite{wang2024auxiliarylossfree}.
Communication operators are represented as nodes annotated with collective type (e.g., All-Reduce), message size, and participating ranks.
The frontend then serializes the central DAG into per-GPU Chakra traces~\cite{sridharan2023chakra} while preserving the graph's intra- and inter-device dependency identifiers. A communication node blocks its dependent chain until completion, while independent ready compute and communication nodes may overlap. Inter-stage dependencies are modeled as point-to-point communication control nodes.

\paragraph{Operator-level performance modeling.}
The hardware-aware compute model then assigns latency to each operator in the graph by considering both compute and memory hierarchy effects.

For GEMMs (attention projections and feed-forward layers), operator shapes reflect the LLM architecture and parallelism configuration. Kernel latency comes from an extended multi-roofline model that constructs a tiled execution timeline.
The model selects the lowest-latency configuration predicted among a catalog of CUTLASS-style tile candidates~\cite{cutlass}. A candidate specifies a thread-block tile (a CUDA cooperative thread array, or CTA), warp tile sizes, software-pipeline depth, and threads per block. Candidates that exceed the target GPU's thread, register, shared-memory, or resident-block limits are discarded and every remaining candidate is priced analytically.

From the chosen tiling the model derives shared-memory, L2 (via a first-touch reuse analysis), and DRAM traffic, along with the number of thread blocks resident per streaming multiprocessor (SM). The grid is divided by this concurrent capacity to obtain the number of execution waves. Each wave's cycle count decomposes into prologue, steady-state, and epilogue phases, with the steady state governed by the slower exposed path: a compute path (shared-memory-to-register movement, tensor-core matrix-multiply-accumulate (MMA) issue, and tensor-core latency across K groups) versus a memory path (L2 and DRAM service cycles feeding the software pipeline). Kernel latency is then
\begin{equation}
t_{\mathrm{GEMM}} \,=\, \frac{N_{\mathrm{full}}\, C_{\mathrm{full}} \,+\, C_{\mathrm{last}}}{f_{\mathrm{clock}}},
\label{eq:extended_roofline}
\end{equation}
where $C_{\mathrm{full}}$ and $C_{\mathrm{last}}$ are the modeled cycles of a full and of the final wave, $N_{\mathrm{full}}$ is the number of full waves, and $f_{\mathrm{clock}}$ is the device clock. Because the dominant path is re-derived for every operator shape, kernel efficiency emerges from the shape and device parameters. The result captures hardware-aware compute-bound and memory-bound regimes without full microarchitectural simulation. Section~\ref{sec:operator_ablation} ablates these model components against NCU-profiled GEMM kernels.

Pointwise and reduction operators (e.g., Softmax) use a simpler analytical bandwidth model based on arithmetic intensity. Fused kernels such as FlashAttention~\cite{flashattn} are treated as composite operators with adjusted tiling and reduced off-chip traffic.

\paragraph{Memory modeling.}
RAPID-LLM checks whether each configuration fits in device memory. It first counts the parameters assigned to each rank after parallelism-driven partitioning, then accounts for optimizer state, gradients, and (for inference) KV caches under the chosen parallelism, precision (mixed vs FP32), and ZeRO~\cite{rajbhandari2020zero} sharding settings. It then traverses the computation graph under the selected recomputation policy (full, selective, or off~\cite{korthikanti2023reducing}), adding and releasing activation tensors as their consumers execute and tracking transient ZeRO-3 parameter materializations. A configuration is feasible only if its persistent state, KV cache, and peak live transient state fit in device memory.

\subsection{Backend: network modeling}

RAPID-LLM executes Chakra \cite{sridharan2023chakra} execution traces on an extended, modified version of Astra-Sim's congestion-aware analytical backend~\cite{astrasim}. The backend constructs explicit multidimensional topologies, models congestion on a per-link basis, and supports soft and hard faults in links, while providing an extended library of base topologies and collective algorithms.

\paragraph{Multidimensional topology construction and congestion.}
RAPID-LLM's backend instantiates all links in each dimension of the network, rather than relying on distance-based abstractions used by congestion-unaware tools such as the original Astra-Sim. Users describe an $N$-dimensional fabric as an ordered list of dimensions, each specifying a base topology, node count, bandwidth, latency, and the TP/CP/EP/PP/DP axes mapped onto it. The backend recursively builds the corresponding structure. A logical rank's parallelism coordinates are flattened in the declared axis order and decoded into one physical coordinate per network dimension, thus partitioning the physical GPU grid. Compared to the simpler networking topology library of Astra-Sim, RAPID-LLM's backend supports additional topologies such as $N$-D Mesh, Torus, KingMesh, and NVIDIA SuperPOD-style fat trees.

\begin{figure*}[!htbp]
    \centering
    \includegraphics[width=\textwidth]{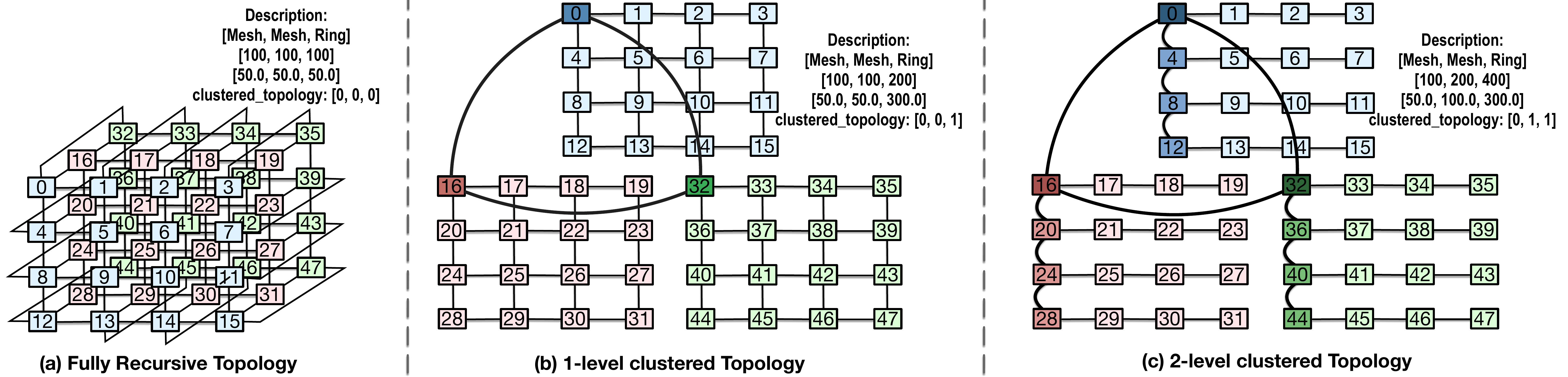}
    \caption{Topology structures supported by RAPID-LLM's network backend. RAPID-LLM is flexible enough to model a fully dense 3D Mesh-Torus hybrid (a), a Ring of connected 2D Meshes (b), and even a Ring of sparsely connected serial GPU chains (c). }
    \Description{Diagrams of the network topology structures supported by the backend, including ring, fully connected, switch-based, mesh, torus, KingMesh, and SuperPOD-style fat-tree fabrics, composed into multidimensional networks.}
    \label{fig:topo_structures}
\end{figure*}

\begin{figure}[!htbp]
    \centering
    \includegraphics[width=0.6\linewidth]{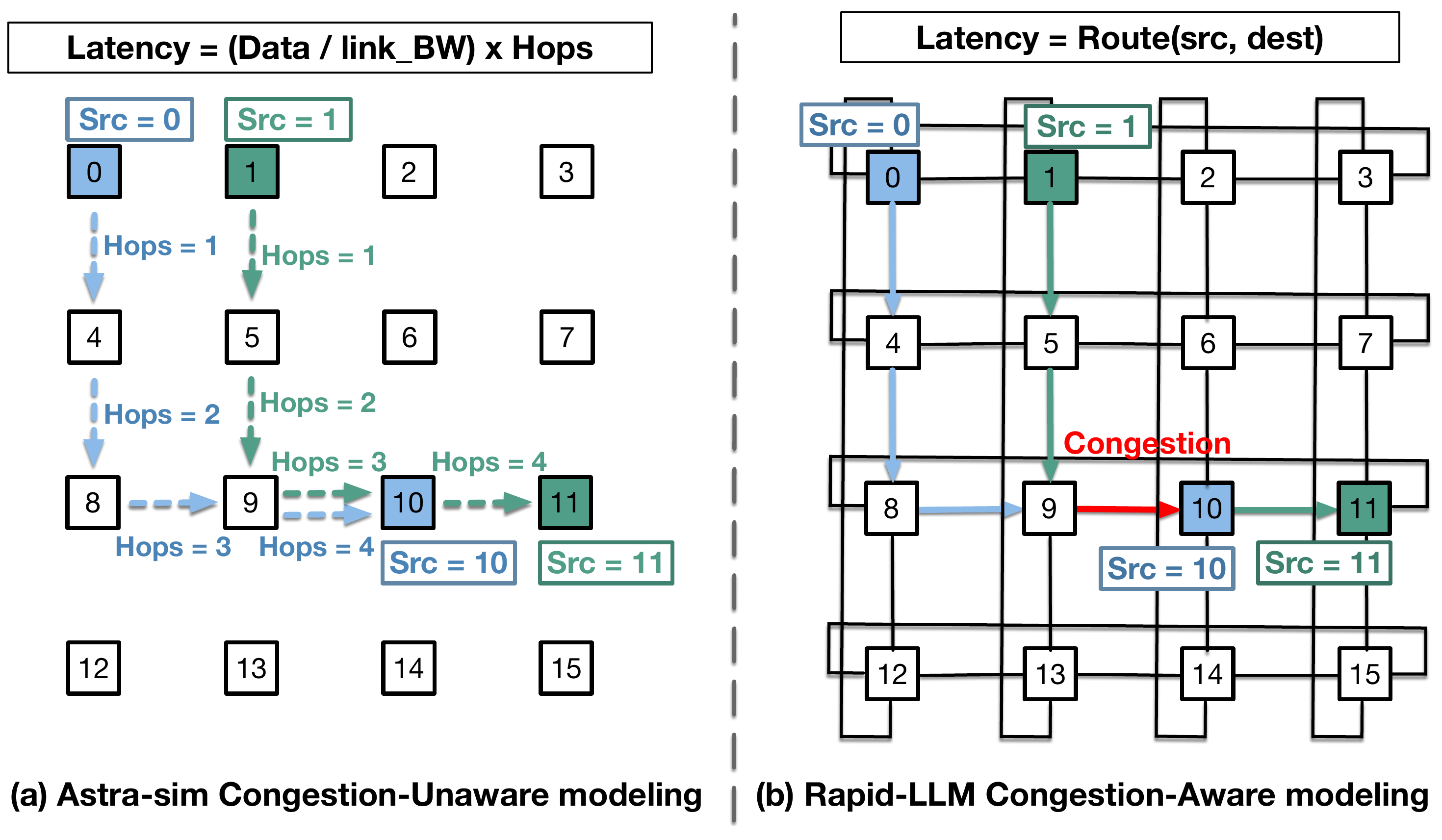}
    \caption{Congestion-aware modeling of multidimensional topologies in RAPID-LLM.}
    \Description{Illustration of a two-dimensional torus where chunks from concurrent collectives share a directed link; per-link first-come-first-served queues serialize the overlapping transfers, modeling congestion.}
    \label{fig:congestion}
\end{figure}

The selected collective algorithm decomposes each collective node into point-to-point chunks, which traverse the links of the mapped dimension in routing order. Every directed link has a first-come-first-served queue. A chunk of $S$ bytes occupies a link of bandwidth $B$ for $S/B$ time units and reaches the next hop after the link latency plus $S/B$ time units. A chunk that arrives while the link is occupied waits in that link's queue. Applying this rule at every hop captures queueing and congestion across dimensions. Figure~\ref{fig:topo_structures} summarizes the supported topology structures, and Figure~\ref{fig:congestion} illustrates shared-link contention.

\begin{figure}[!htbp]
    \centering
    \includegraphics[width=0.6\linewidth]{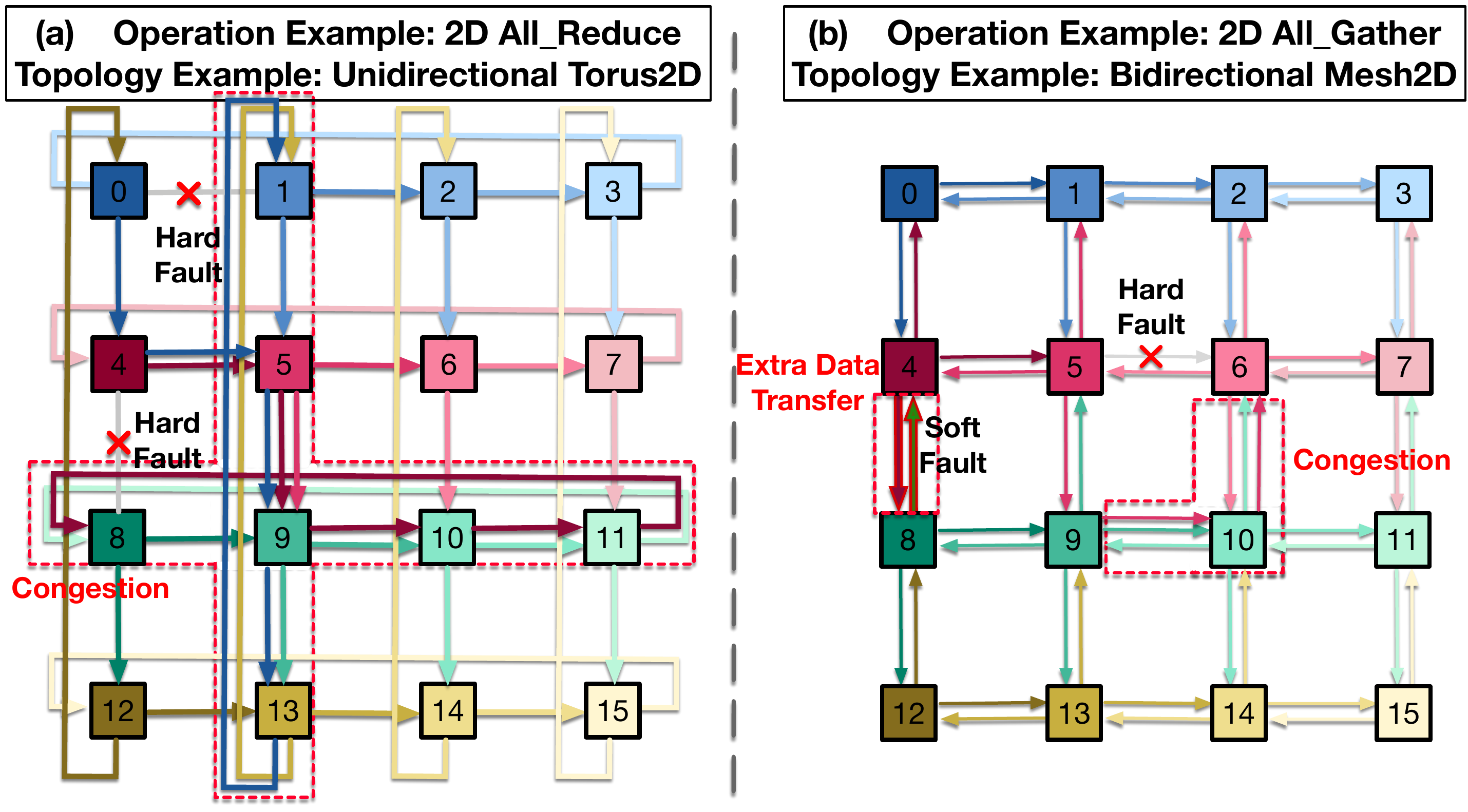}
    \caption{Examples of fault presence and RAPID-LLM's routing to avoid hard faults, along with the resulting congestion on alternate links.}
    \Description{Illustration of hard link faults removing links from a topology, with fault-aware routing redirecting traffic onto alternate healthy links and creating additional congestion on those links.}
    \label{fig:fault}
\end{figure}

\paragraph{Fault-aware network modeling.} RAPID-LLM's backend models both soft faults, where a link remains usable but at reduced effective bandwidth or higher latency (e.g., due to retries), and hard faults, where a link is removed from the topology. Given a fault configuration (pairs of endpoints and fault rates), the backend adjusts per-link bandwidth, and routes collectives using fault-aware variants of dimension-order routing that avoid unusable links when possible. Traffic rerouted around faults may break symmetry and can create new congestion hotspots that our backend captures. Figure~\ref{fig:fault} shows two examples where hard faults force traffic onto alternate paths and induce additional congestion. 

\subsection{Scalability}
A typical networking backend like AstraSim consumes one execution graph per device, so a materialized trace grows with the product of GPU count, layer depth, microbatches, and iterations~\cite{sridharan2023chakra,astrasim}. RAPID-LLM avoids forming this product by exploiting the homogeneity of modern LLMs. A transformer is a stack of identical layers replayed across microbatches and iterations, and large-scale deployments map that stack homogeneously, so every tensor/context-parallel group has the same operator shapes and every pipeline stage runs the same schedule. The frontend therefore emits a small set of distinct graphs: the forward and backward pass of a single transformer layer, soft-or-hard faulted variants of each, and one pipeline graph across stages and microbatches. These are reused by the backend, reducing graph growth across repeated layers, microbatches, and equivalent device groups.

Reuse is valid when the collectives repeated within a layer (TP, CP, and EP) occupy equivalent inner-network positions while PP and DP span the outer dimensions, so that replayed instances encounter the same congestion. Cases with link faults cannot use this optimization, as link faults break symmetry across device groups.

This optimization is required to make execution of >1000 GPU cases feasible. For example, for the A100 training cases of Section~\ref{sec:a100_train}, it reduces the number of network events fed to the backend by 129$\times$--145$\times$ relative to a fully-flattened per-device trace execution.

\section{Validation}
\label{sec:validation}
This section evaluates how well RAPID-LLM predicts end-to-end time per training batch and inference step latency for representative LLM workloads.

\subsection{Methodology}

\paragraph{Reference sources.}
Ground truth comes from four sets of published measurements, spanning A100 and H100 devices across dense, fully sharded, and mixture-of-experts (MoE) regimes: Llama inference on both devices~\cite{nvidia,nvidiaNIM}, GPT-scale training on A100 clusters~\cite{kundu2024altnewdeepflow}, MosaicML's LLM-foundry MPT benchmarks on H100~\cite{mosaicml_llmfoundry_mpt_benchmarks}, and Megatron-Core MoE training on H100~\cite{liu2025moefolding}. The network backend is validated separately against ns-3 (Section~\ref{sec:ns3}). Where a comparison tool can express the same case, we also report GenZ~\cite{genz}, Vidur~\cite{vidur}, llm-analysis~\cite{li2023llmanalysis}, and STAGE~\cite{man2025stage} alongside RAPID-LLM.

\paragraph{Hardware configuration.}
Per-GEMM kernel latency comes from RAPID-LLM's extended-roofline operator model. Each device then uses three calibrated constants shared across all cases: residual compute utilization ($85\%$ for H100 and $100\%$ for A100), HBM utilization of $80\%$, network utilization of $80\%$.

The compute utilization asymmetry is caused by effects that are out of scope for analytical design space exploration tools, such as thermal throttling~\cite{go2025characterizing,latif2025cooling}, which can be significant considering the H100's larger TDP (700\,W vs 400\,W)~\cite{nvidiah100datasheet,nvidiaa100datasheet}.

The factors were fitted on a stratified 50:50 calibration/holdout split of a subset of the dense inference and training cases and held fixed thereafter (held-out vs calibration MAPE differences remained $\leq2\%$ throughout). The H100 fully sharded and mixture-of-experts training suites were not used for calibration purposes, so in total only 31 out of the 124 total validation cases were used for calibration.

All MoE validation cases assume balanced token routing across experts.

\paragraph{Network configuration.}
Every case is simulated end-to-end assuming networking at physical per-GPU NVLink rates~\cite{nvidiah100datasheet,nvidiaa100datasheet} and, whenever the number of GPUs exceeds a single DGX Box, we assume NVIDIA DGX SuperPOD fabric rates~\cite{nvidiadgxsuperpod}.

\subsection{Inference validation}
\begin{figure*}[!t]
    \centering
    \includegraphics[width=\textwidth]{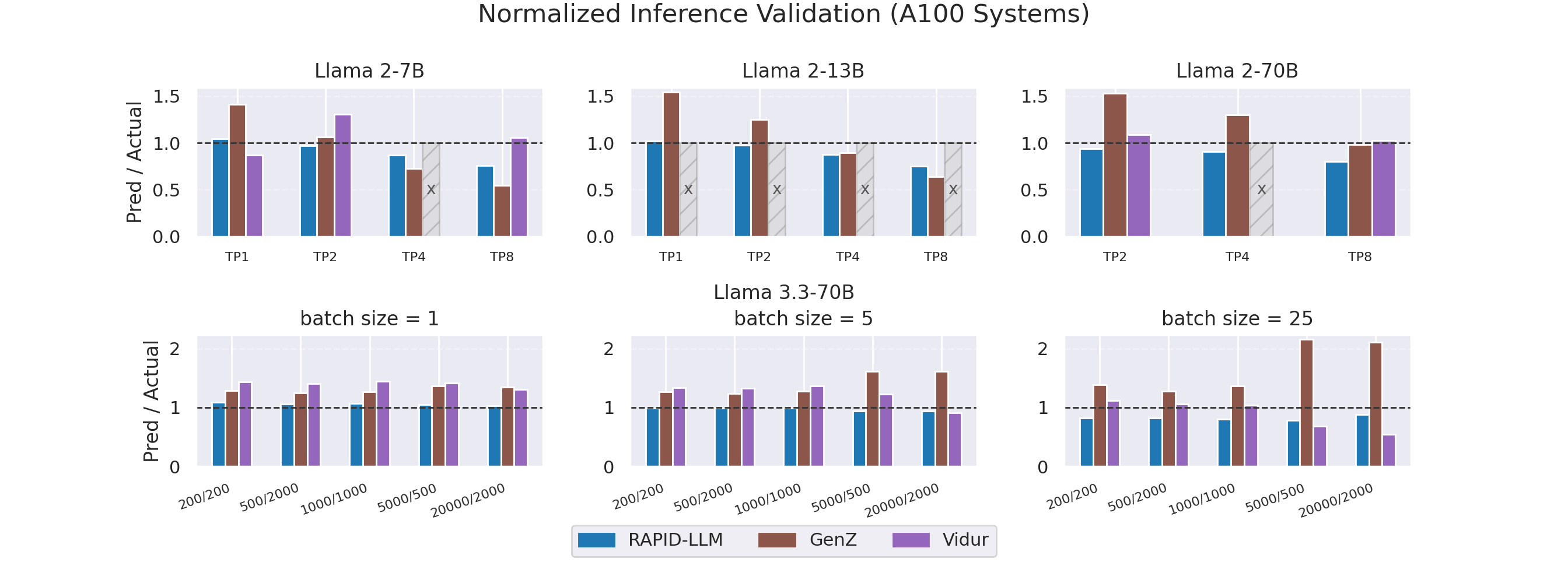}\\[2pt]
    \includegraphics[width=\textwidth]{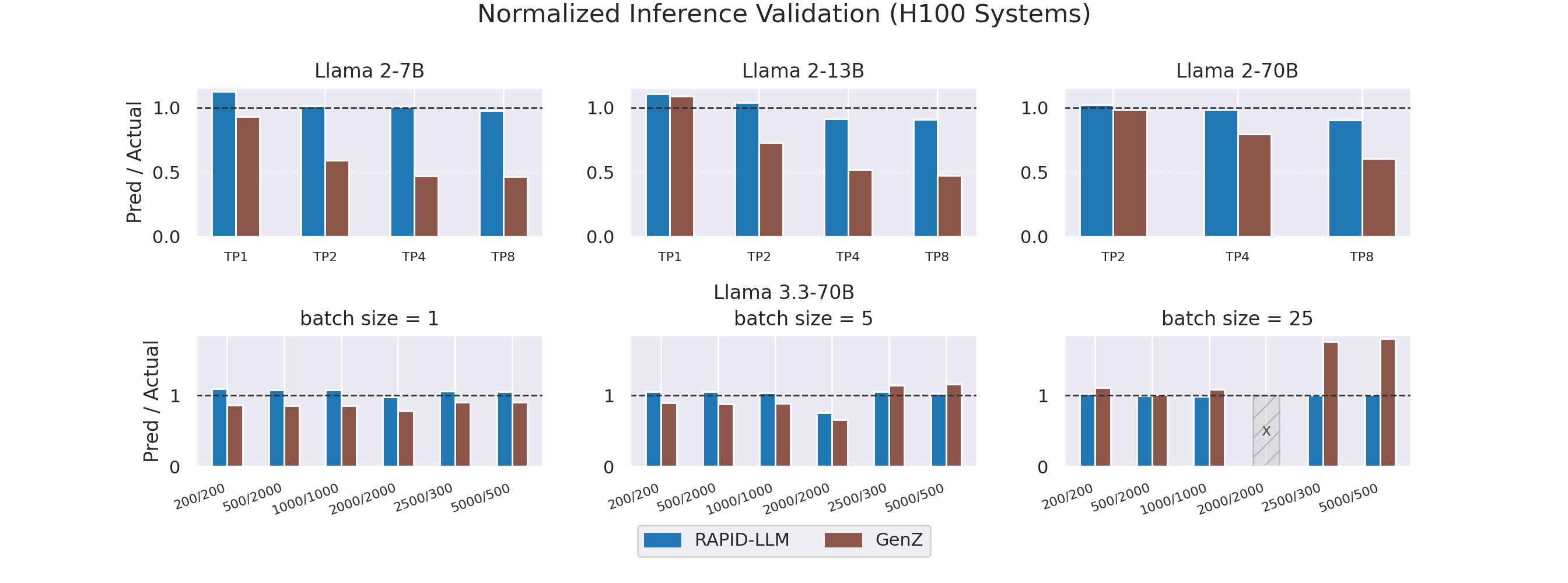}
    \caption{Normalized inference validation on A100 (top) and H100 (bottom) SXM systems~\cite{nvidia,nvidiaNIM}. Within each grid, the upper row shows Llama2 models under pure tensor parallelism (batch size = 1, input 200 / output 200) and the lower row shows Llama3.3-70B across input/output token lengths and batch sizes. MAPE is 9.8\% on A100 and 5.4\% on H100. Hatched cells mark configurations without published reference data.}
    \Description{Two grids of normalized inference latency bars comparing RAPID-LLM, GenZ, and Vidur predictions against measured A100 and H100 latencies for Llama2 tensor-parallelism sweeps and Llama3.3-70B token-length and batch-size cases; hatched cells mark configurations without published reference data.}
    \label{fig:nvidia_inf}
\end{figure*}

Figure~\ref{fig:nvidia_inf} evaluates total inference latency of modern Llama models under pure tensor parallelism (TP) across a range of TP degrees. These cases stress operator shape changes (prefill-like GEMMs versus decode-like skinny GEMMs) and model sensitivity to TP communication frequency. Here, input token length denotes prompt length (prefill), and output token length denotes generated length (decode).

\subsubsection{A100 SXM systems}
We compare RAPID-LLM against published measurements~\cite{nvidia,nvidiaNIM} alongside state-of-the-art inference serving prediction tools, GenZ \cite{genz} and Vidur \cite{vidur}. While RAPID-LLM and GenZ are flexible enough to model all cases, Vidur's repo did not natively provide A100 profiling data for several cases (e.g., Llama2-13B), leading to its omission in those comparisons. Across the tensor-parallelism sweep and all input/output-length and batch-size pairs, RAPID-LLM achieves a mean absolute percentage error (MAPE) of 9.8\%, against 38.9\% for GenZ and 24.7\% for Vidur over the 20 cases its profiles cover.

\subsubsection{H100 SXM systems}
The bottom grid of Figure~\ref{fig:nvidia_inf} repeats the study on H100 SXM systems, covering the same Llama2 tensor-parallelism sweep and the NIM Llama3.3-70B cases. Across the 28 H100 cases RAPID-LLM reaches a MAPE of 5.4\%, against 25.3\% for GenZ.

\subsection{Training validation}

\begin{figure}[!htbp]
    \centering
    \includegraphics[width=1\linewidth]{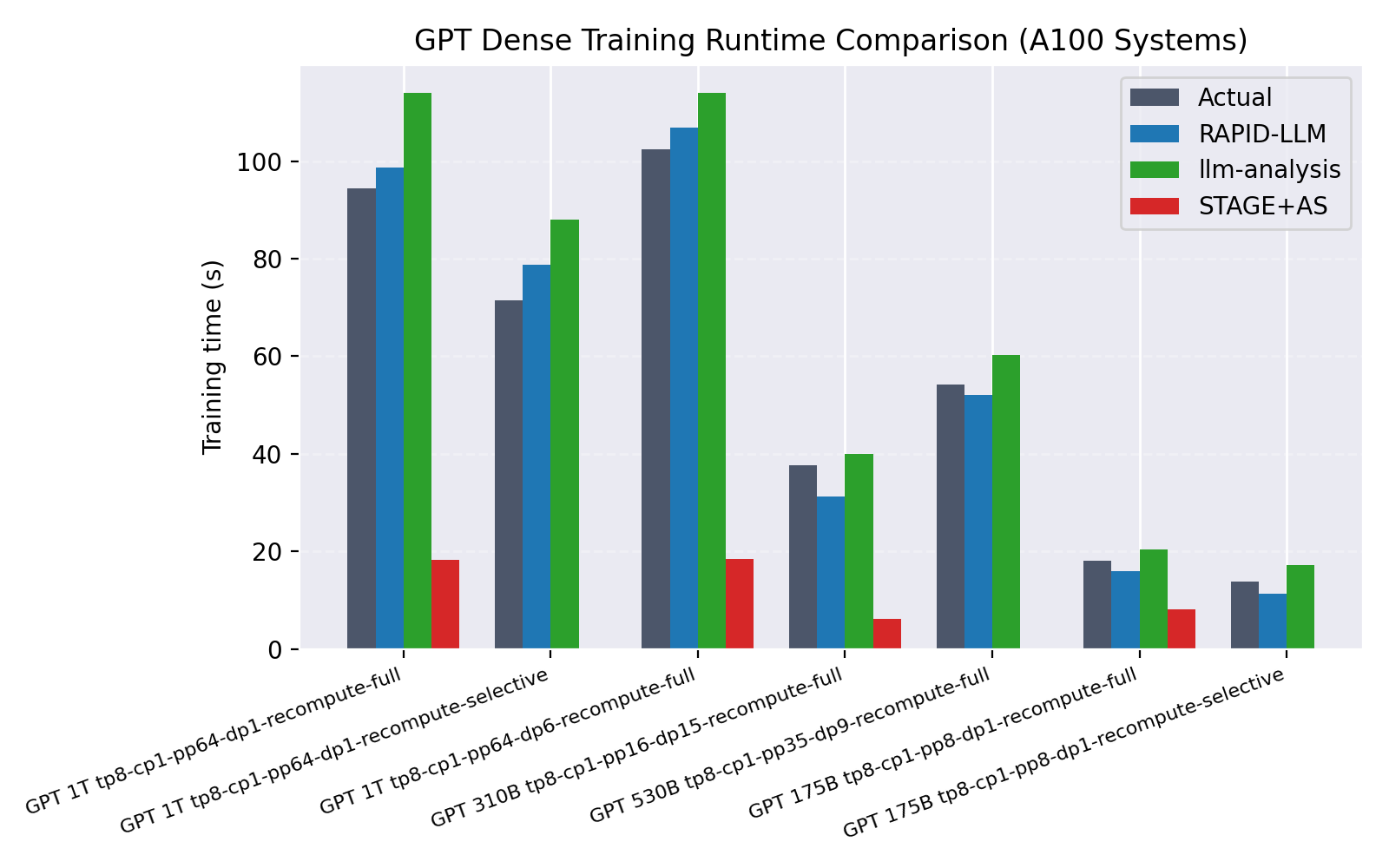}
    \caption{Predicted training time per batch across GPT-scale configurations on A100 clusters, compared to published reference runtimes \cite{kundu2024altnewdeepflow}, llm-analysis \cite{li2023llmanalysis}, and STAGE paired with Astra-Sim \cite{man2025stage,astrasim}. We distinguish selective versus full activation recomputation regimes. RAPID-LLM reaches a MAPE of $9.9\%$ (llm-analysis: $15.8\%$). Missing STAGE bars are configurations it could not complete.}
    \Description{Grouped bar chart of predicted training time per batch from RAPID-LLM, llm-analysis, and STAGE against published reference runtimes across GPT-scale A100 configurations from 64 to 3072 GPUs, split into selective and full activation recomputation regimes.}
    \label{fig:nvidia_train}
\end{figure}

\subsubsection{Dense training on A100}
\label{sec:a100_train}
Figure~\ref{fig:nvidia_train} compares RAPID-LLM to published runtimes, llm-analysis~\cite{li2023llmanalysis}, and the symbolic-trace generator STAGE~\cite{man2025stage} paired with Astra-Sim on large-scale training configurations ranging from 64 to 3072 GPUs. We run llm-analysis at its own documented settings and report both its raw output and the correction described in Appendix~\ref{app:train_comparison}. The raw output reaches 68.2\% MAPE. After correcting the ZeRO-3-style weight-All-Gather floor that it applies to these unsharded-DP cases, it reaches 15.8\% MAPE, while RAPID-LLM reaches 9.9\%. A fixed global efficiency cannot track how kernel efficiency shifts across operator shapes and recomputation regimes, which is precisely what RAPID-LLM's shape-aware operator model captures. On the four configurations STAGE completes, it underestimates runtimes by 55--84\%, and its activation-recomputation option does not alter the generated traces, which explains the discrepancy (Appendix~\ref{app:train_comparison}).

\subsubsection{Fully sharded training}
\begin{figure}[!htbp]
    \centering
    \includegraphics[width=0.85\linewidth]{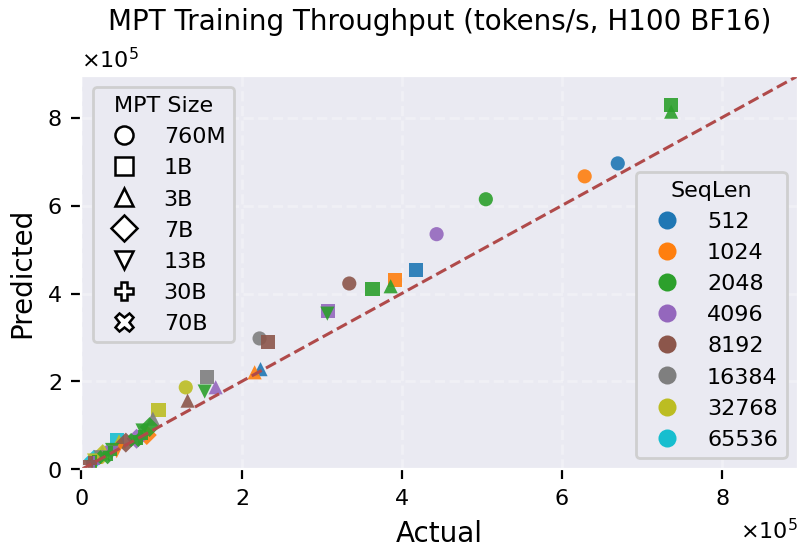}
    \caption{Predicted vs. published training throughput for MosaicML MPT models (760M--70B) on H100 BF16 \cite{mosaicml_llmfoundry_mpt_benchmarks}, across sequence lengths 512--65k and 8--64 GPUs under FSDP with gradient accumulation. MAPE across all 52 configurations is $13.2\%$.}
    \Description{Parity scatter plot of predicted versus published training throughput for MosaicML MPT models on H100 under FSDP, with points for all 52 configurations lying close to the diagonal.}
    \label{fig:mpt_train}
\end{figure}

We validate against MosaicML's published LLM-foundry MPT training benchmarks on H100 \cite{mosaicml_llmfoundry_mpt_benchmarks}. Each run shards parameters, gradients, and optimizer state across all data-parallel ranks (FSDP, equivalent to ZeRO-3), which adds a parameter all-gather and a gradient reduce-scatter on every step and stresses the network. Across all 52 rows, spanning MPT 760M to 70B, sequence lengths 512 to 65k, and 8 to 64 GPUs, RAPID-LLM reaches a MAPE of 13.2\% (Figure~\ref{fig:mpt_train}).

\subsubsection{Mixture-of-experts training}
\begin{figure}[!htbp]
    \centering
    \includegraphics[width=1\linewidth]{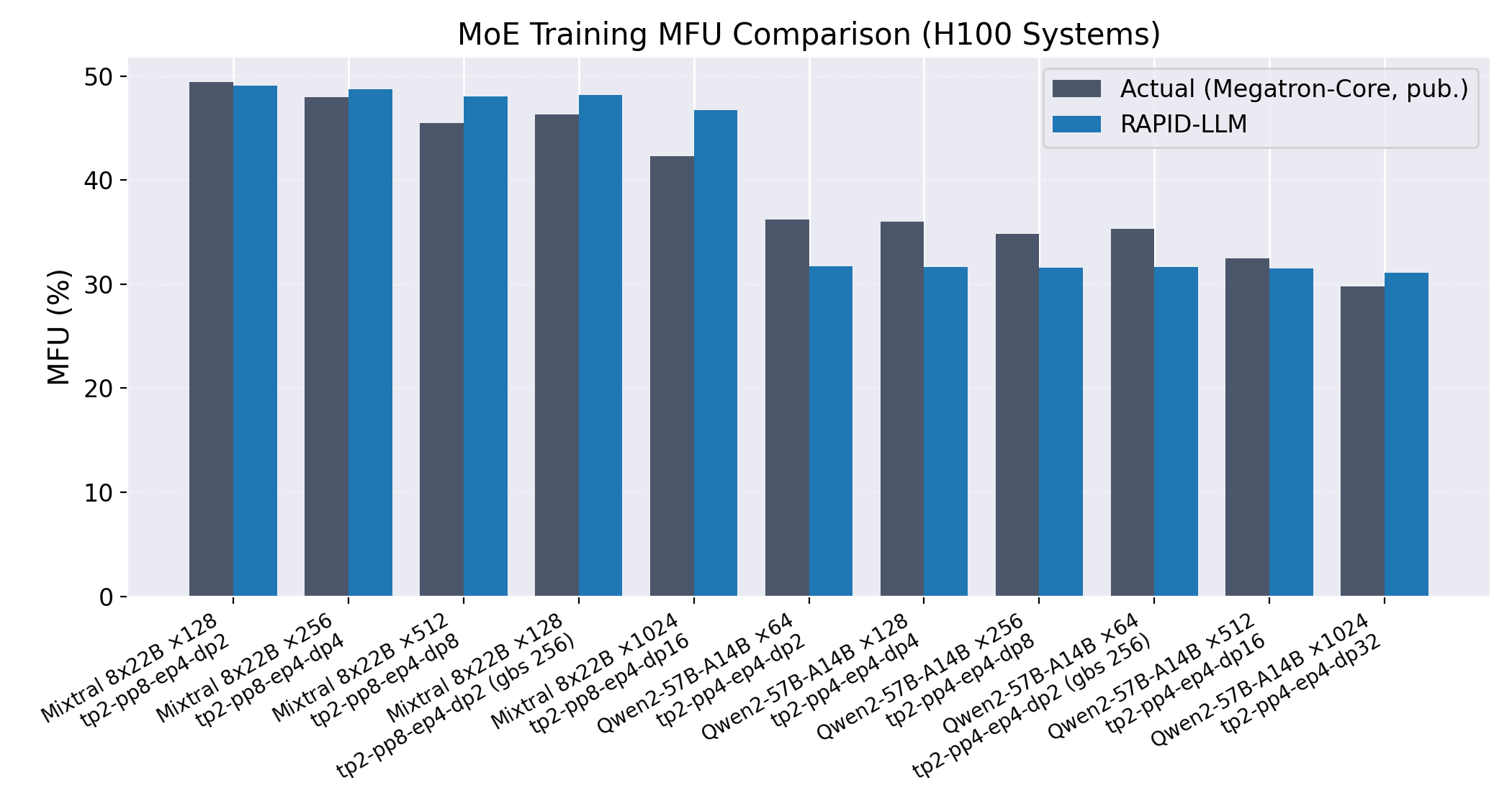}
    \caption{Predicted MFU for Mixtral 8x22B and Qwen2-57B-A14B MoE training on H100 clusters (64--1024 GPUs, combined TP/PP/EP/DP), compared to published Megatron-Core MFU \cite{liu2025moefolding}. MAPE is $6.7\%$.}
    \Description{Bar chart comparing RAPID-LLM predicted MFU with published Megatron-Core MFU for Mixtral 8x22B and Qwen2-57B-A14B mixture-of-experts training across 64 to 1024 H100 GPUs.}
    \label{fig:moe_train}
\end{figure}

A mixture-of-experts layer routes each token to a subset of experts, adding a pair of all-to-all exchanges per layer that the dense suites above never exercise. Figure~\ref{fig:moe_train} evaluates this regime against published Megatron-Core measurements on H100 clusters \cite{liu2025moefolding} for two representative MoE architectures: Mixtral 8x22B~\cite{mistral2024mixtral8x22b,jiang2024mixtral} (coarse-grained, 8 experts, top-2) and Qwen2-57B-A14B~\cite{yang2024qwen2} (fine-grained, 64 experts, top-8, with shared experts), spanning 64 to 1024 GPUs under combined TP/PP/EP/DP mappings. We compare in the MFU basis the source publishes, converting RAPID-LLM's simulated iteration times to MFU. Across the 11 MoE configurations RAPID-LLM predicts MFU with a MAPE of 6.7\%.

Taken together, these validation results demonstrate that RAPID-LLM is both accurate and scalable across realistic dense and mixture-of-experts training configurations.

\subsection{Component-level validation}
Beyond the end-to-end suites above, we also validate RAPID-LLM's two main modeling components in isolation: the frontend's GEMM operator model against profiled GPU kernels, and the network backend against packet-level simulation.

\subsubsection{Operator-model ablation}
\label{sec:operator_ablation}
\begin{figure}[!htbp]
  \centering
  \includegraphics[width=\linewidth]{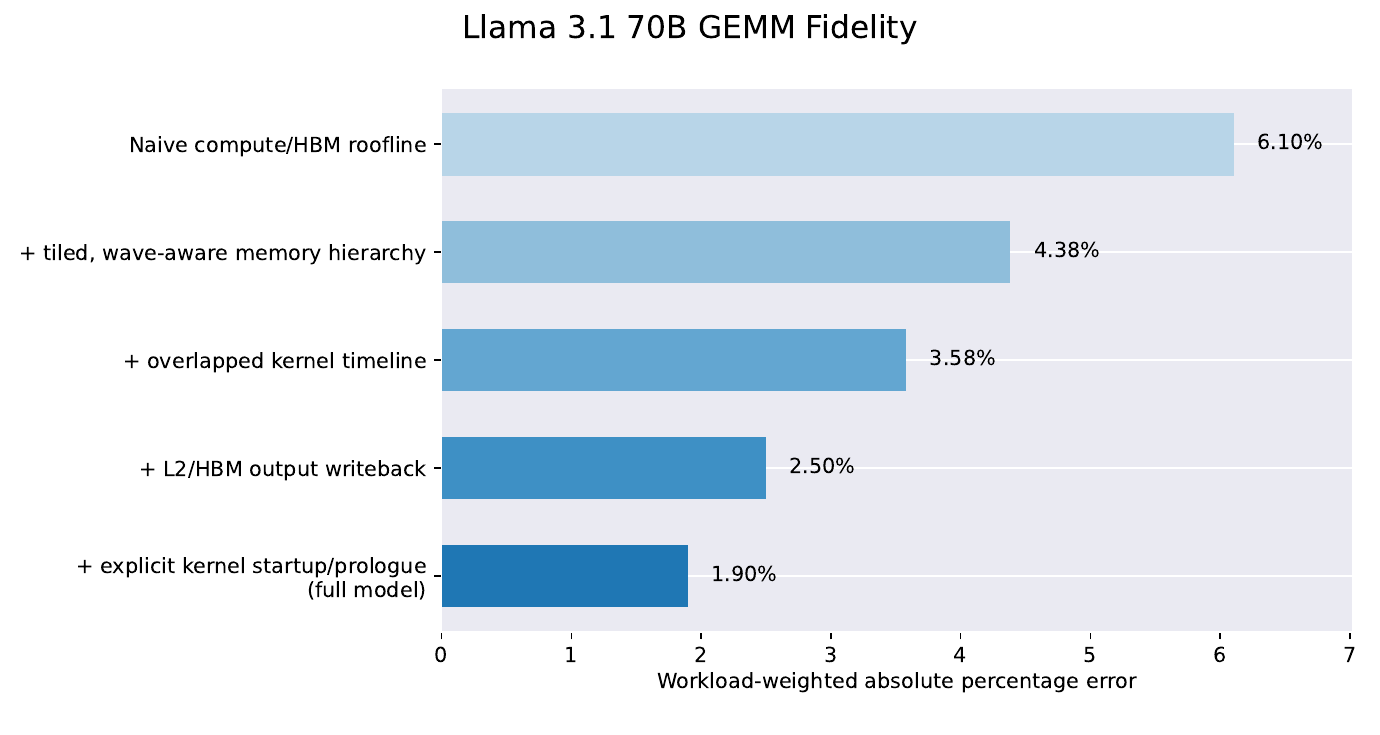}
  \caption{Cumulative GEMM-model ablation for a Llama3.1-70B forward pass on A100, weighted over 321 projection, FFN, and LM-head GEMM calls.}
  \Description{Bar chart showing workload-weighted absolute percentage error on profiled A100 GEMMs decreasing from 6.1 percent for a naive roofline model to 1.9 percent as cumulative components of the extended operator model are enabled.}
  \label{fig:gemm-operator-ablation}
\end{figure}

Figure~\ref{fig:gemm-operator-ablation} evaluates the cumulative model components on A100 NCU-profiled GEMM shapes from a 1,024-token Llama3.1-70B forward pass, weighted by their execution time. This is a GEMM-only comparison that neglects networking/parallelism effects, and focuses only on raw kernel execution time. The operator-level modeling of RAPID-LLM reduces MAPE from 6.1\% to 1.9\% by accurately tracking GPU kernel performance for different GEMM shapes.

\subsubsection{Network backend validation against ns-3}
\label{sec:ns3}
\begin{figure}[!htbp]
    \centering
\includegraphics[width=\linewidth]{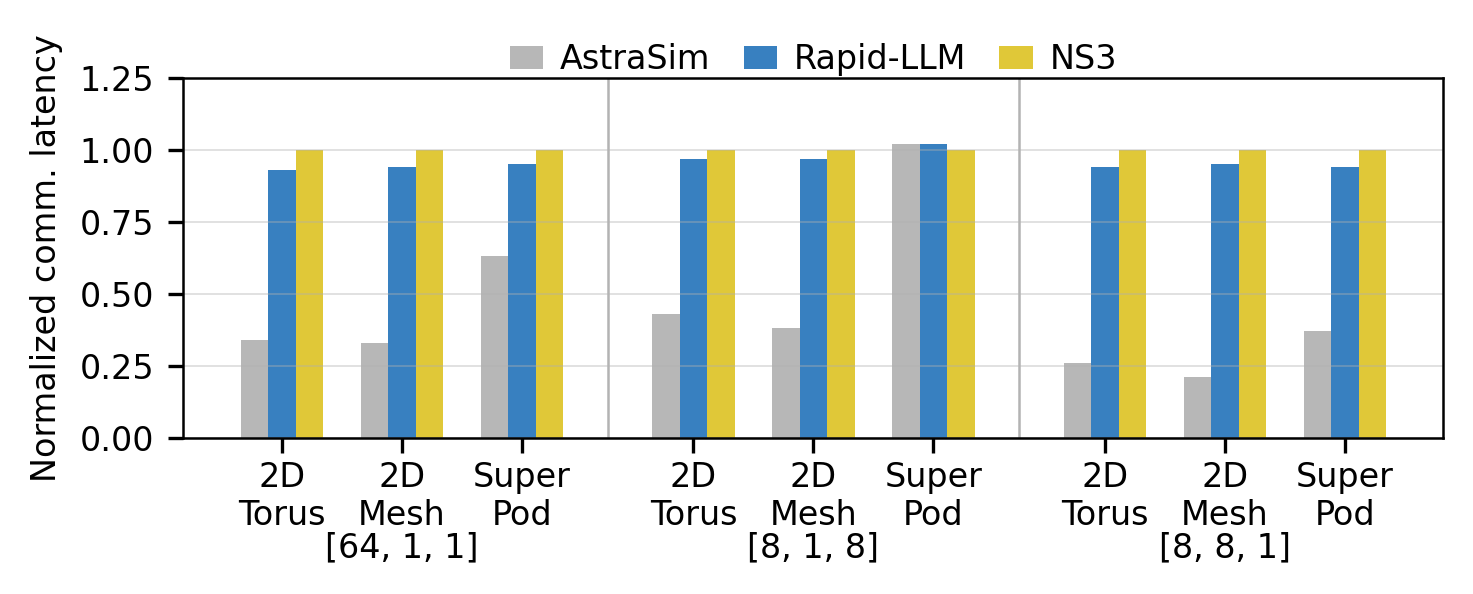}
    \caption{Network backend validation against Astra-Sim (stock) and ns-3 for three communication workloads corresponding to different parallelism patterns [dp,tp,pp] for Llama3.1-70B (sequence length 4096, batch size 128). RAPID-LLM stays within 8\% of ns-3 across the shown topologies.}
    \Description{Bar chart comparing communication time predicted by stock Astra-Sim, RAPID-LLM, and ns-3 for three parallelism-induced communication workloads on different topologies; RAPID-LLM stays within 8 percent of ns-3 in all cases.}
    \label{fig:ns3_validation}
\end{figure}

We validate the network backend by comparing its predicted communication time against ns-3, a widely used discrete-event packet-level network simulator \cite{ns3}. We focus on communication patterns induced by different parallelism axes, since these stress different collective and point-to-point behaviors. For each topology, all evaluations are performed using a collective communication algorithm specifically tailored to that topology.
Figure~\ref{fig:ns3_validation} compares three backends: Astra-Sim (stock), RAPID-LLM's extended backend, and ns-3, across the topologies shown in the figure.
The workload corresponds to the communication structure of Llama3.1-70B training on an A100 80GB-based system.

Across all tested cases, RAPID-LLM stays within 8\% of ns-3. For congestion-free communication patterns such as pipeline-parallel point-to-point transfers, Astra-Sim (stock) and RAPID-LLM closely match each other, as expected. For cases with multi-flow contention, RAPID-LLM tracks ns-3 more closely due to explicit per-link queueing and multidimensional contention modeling.
We use ns-3 only for small validation cases because packet-level simulation becomes expensive as network size and the number of concurrent flows increase. Furthermore, to keep execution time manageable, the reported runtime does not consider any compute nodes and only consists of the aggregated collectives that the training runs would exercise.

\section{Case Studies}
\label{sec:case_studies}
RAPID-LLM targets questions that are hard to answer with trace replay or coarse, layer-level models: how sensitive performance is to hybrid-parallel mapping on a fixed cluster, how link faults change end-to-end time, how topology choices affect runtime for different workloads, and how future GPU designs or packaging changes would affect end-to-end performance. This section presents four short studies along those axes.


\subsection{SuperPOD scaling and network sensitivity across parallelism configurations}
\begin{figure}[!htbp]
    \centering
    \includegraphics[width=1\linewidth]{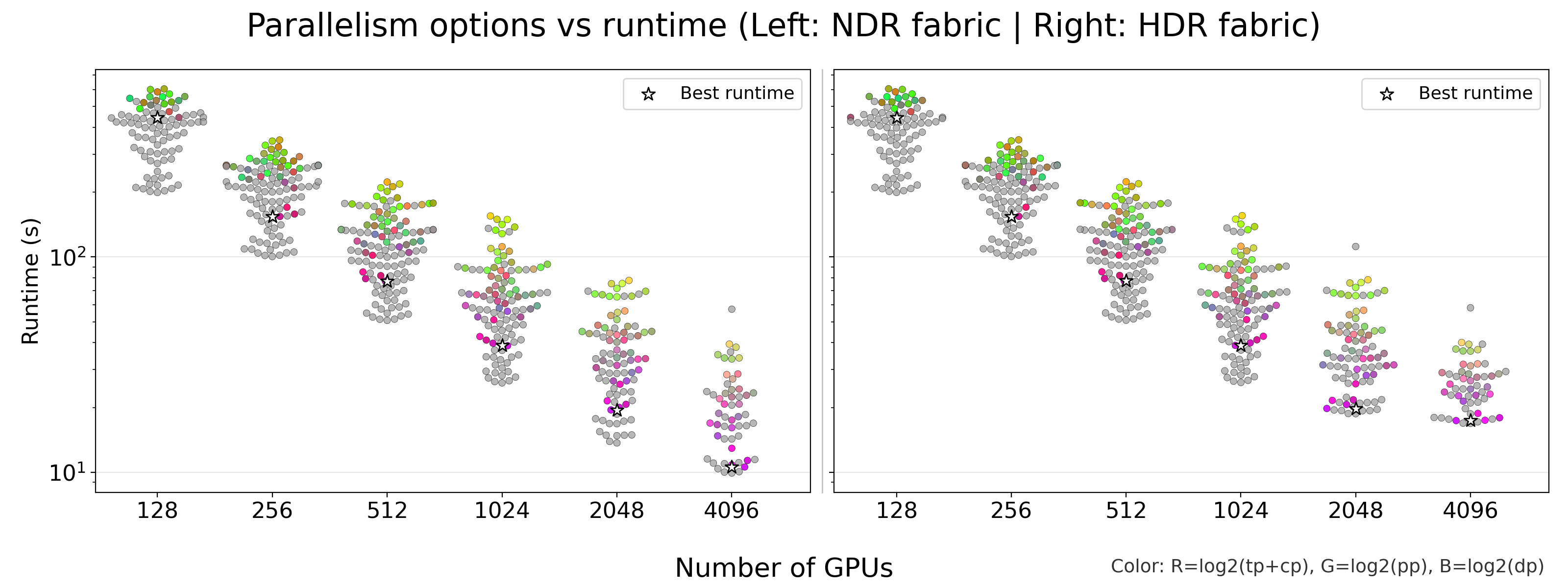}
    \caption{Hybrid-parallel sweep for A100 80GB Llama3.1-405B training (global batch size of 64, sequence length of 33k tokens) on a 2-level SuperPOD-style cluster, comparing two inter-node fabric generations: H100-SuperPOD-class NDR provisioning (400~GB/s per node, left) and A100-SuperPOD-class HDR provisioning (200~GB/s per node, right). Each point is one TP/CP/PP/DP configuration at constant total work. Gray points exceed device memory and are rejected, while other points are colored according to their parallelism type. Stars mark the best configuration per GPU count.}
    \Description{Scatter plots of training runtime versus GPU count for enumerated hybrid-parallel configurations on NDR and HDR SuperPOD fabrics; gray points are memory-infeasible, colored points encode parallelism type, and stars mark the best configuration per GPU count.}
    \label{fig:pod_sweep}
\end{figure}

For a Llama3.1-405B training baseline with fixed batch size, we systematically enumerate integer DP/TP/PP/CP factorizations over 2-level-SuperPOD-style clusters ranging from 128 to 4096 GPUs, evaluating 1,121 mapping--fabric points and pruning configurations that violate memory capacity. Figure~\ref{fig:pod_sweep} compares two generations of inter-node fabric on the same cluster: H100-SuperPOD-class NDR provisioning (400~GB/s per node, left) and A100-SuperPOD-class HDR provisioning (200~GB/s per node, right).

On the NDR fabric the best mapping scales near-ideally through 4096 GPUs. The HDR fabric is indistinguishable up to 2048 GPUs but becomes the binding resource at 4096, where its best mapping runs 1.64$\times$ slower than on NDR. RAPID-LLM thus identifies both when a fabric generation becomes the binding resource for scale-out and which mappings (data-parallel-wide and pipeline-light) fall off the cliff first.

\subsection{Faulty links and fault-aware parallelism selection}

\begin{figure}[!htbp]
    \centering
    \includegraphics[width=\linewidth]{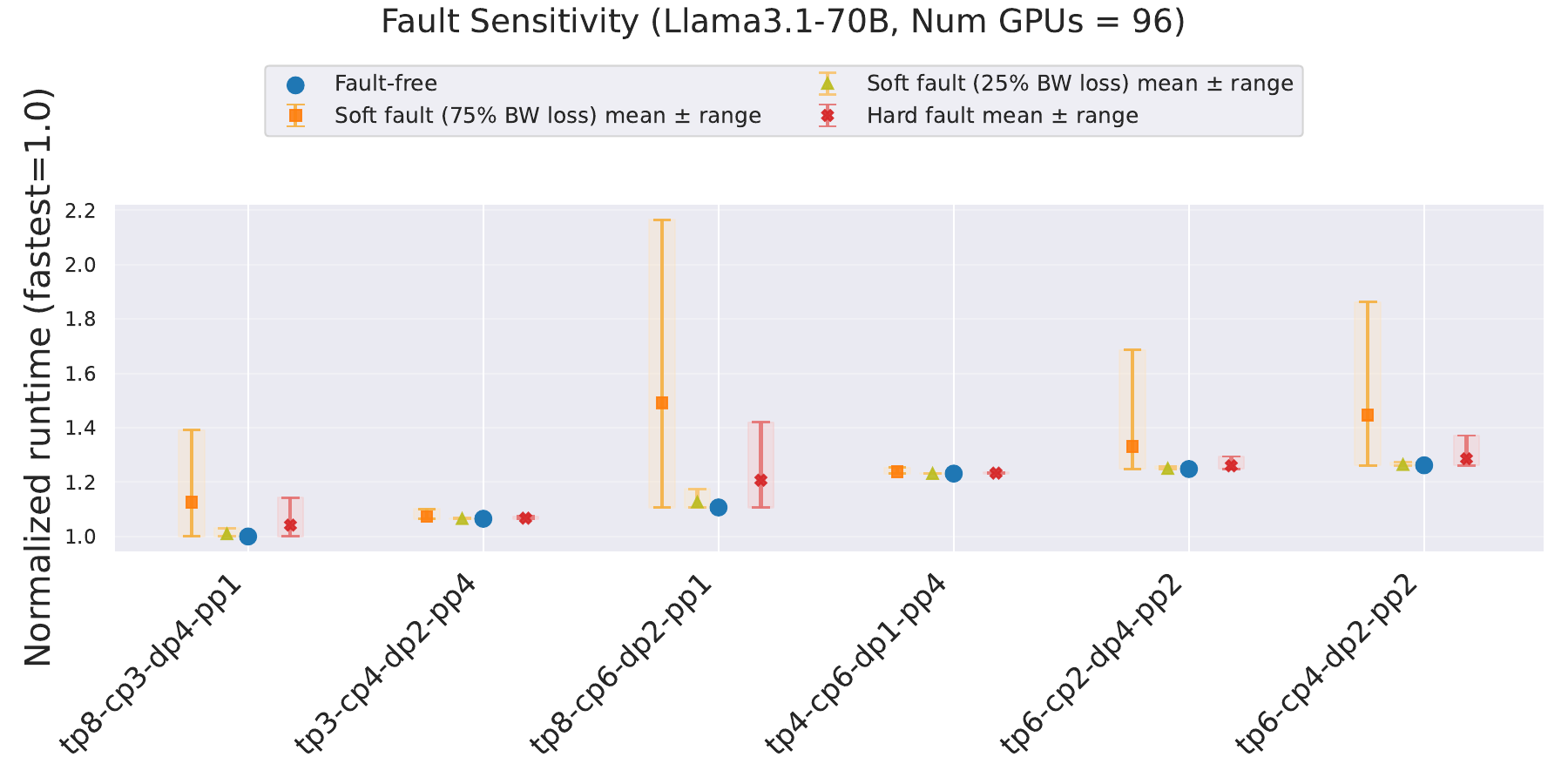}
    \caption{Fault sensitivity across six memory-feasible hybrid-parallel configurations for Llama3.1-70B training on 96 A100 GPUs. Soft faults remove 75\% or 25\% of link bandwidth. Hard faults remove the affected link and trigger rerouting. Markers report the exhaustive means. Bars show the full range over link locations.}
    \Description{Chart of training slowdown for six hybrid-parallel configurations under soft faults that remove 75 or 25 percent of one link's bandwidth and hard faults that remove the link; markers show exhaustive means and bars show the full range over link locations.}
    \label{fig:fault_llama}
\end{figure}

Next we study how degraded and failed links translate into end-to-end training slowdown, and how that slowdown depends on the hybrid-parallel layout. The base setup is Llama3.1-70B training on 96 A100 80~GB GPUs, with a 2D Torus for TP/CP and a linear-chain fabric (1D Mesh) for PP/DP. For each memory-feasible configuration studied, we exhaustively place one fault at each link location in the first network dimension.

A soft fault leaves the link active but reduces its bandwidth by either 75\% or 25\%. A hard fault removes the affected link, causing fault-aware routing to redirect traffic over healthy links. Runtime sensitivity depends on the parallelism layout: configurations that rely heavily on TP and CP (frequent, heavy collectives) tend to be more brittle than configurations that shift sharding toward pipeline parallelism (less frequent, small point-to-point transfers).

Figure~\ref{fig:fault_llama} shows that this dependence is strong. Across the six configurations, a 75\% bandwidth loss produces mean slowdowns of up to 34.8\%, while a 25\% loss produces only up to 1.8\%. Hard faults produce mean slowdowns of up to 9.1\%. Under our routing assumptions, severe soft faults can hurt more than hard faults: traffic remains pinned to a degraded link, while a failed link is removed and its traffic is rerouted over healthy alternatives.

More importantly, severe degradation can reverse the ordering of otherwise competitive configurations. Under 75\% bandwidth loss, the fault-free optimum, \texttt{tp8-cp3-dp4-pp1} becomes 4.9\% slower than \texttt{tp3-cp4-dp2-pp4} in the median case. Parallelism selection based only on fault-free performance can therefore choose a brittle mapping, while fault severity and routing response can materially change which configuration performs best.

\subsection{Topology choice on a waferscale-style system}
\begin{figure}[!htbp]
    \centering
    \includegraphics[width=0.49\linewidth]{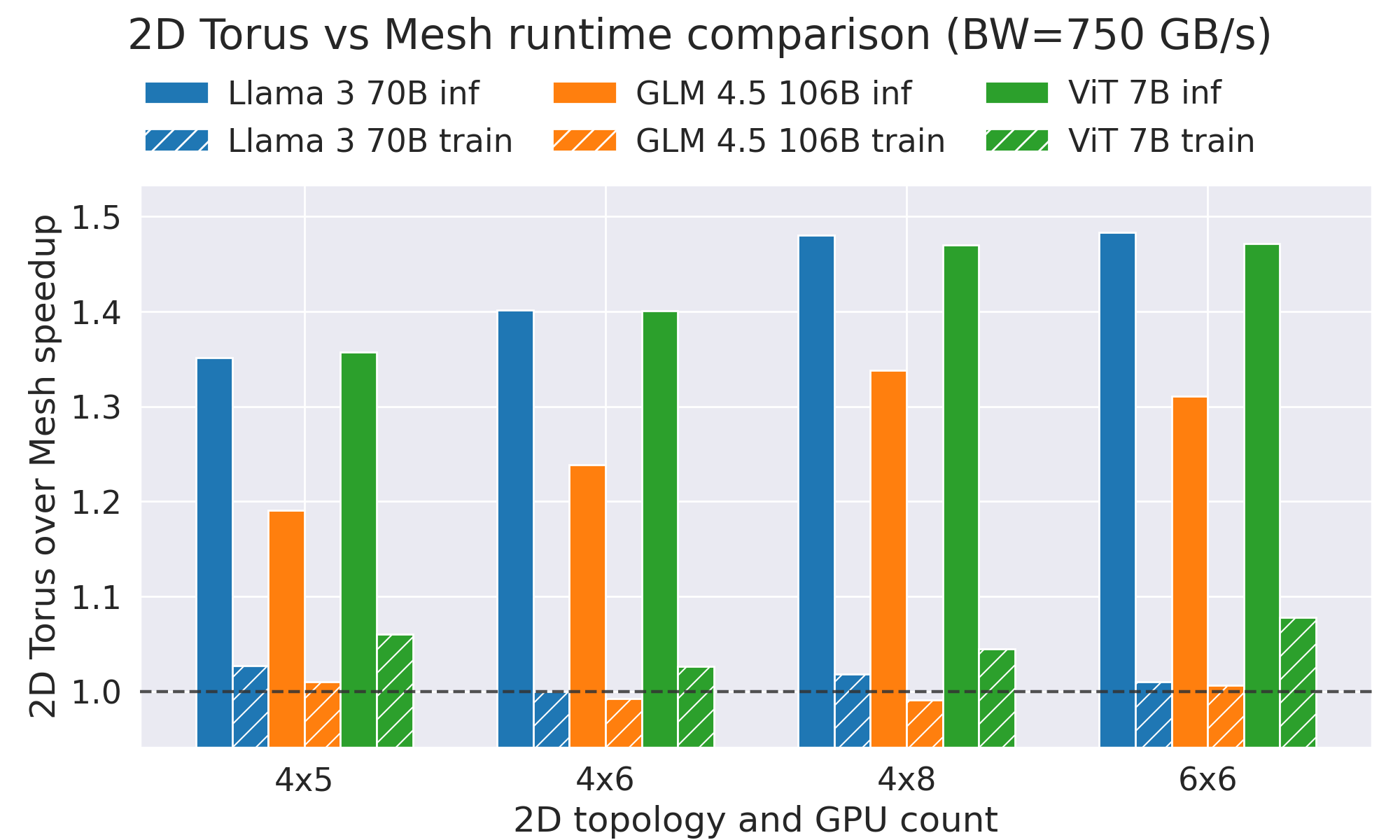}
    \hfill
    \includegraphics[width=0.49\linewidth]{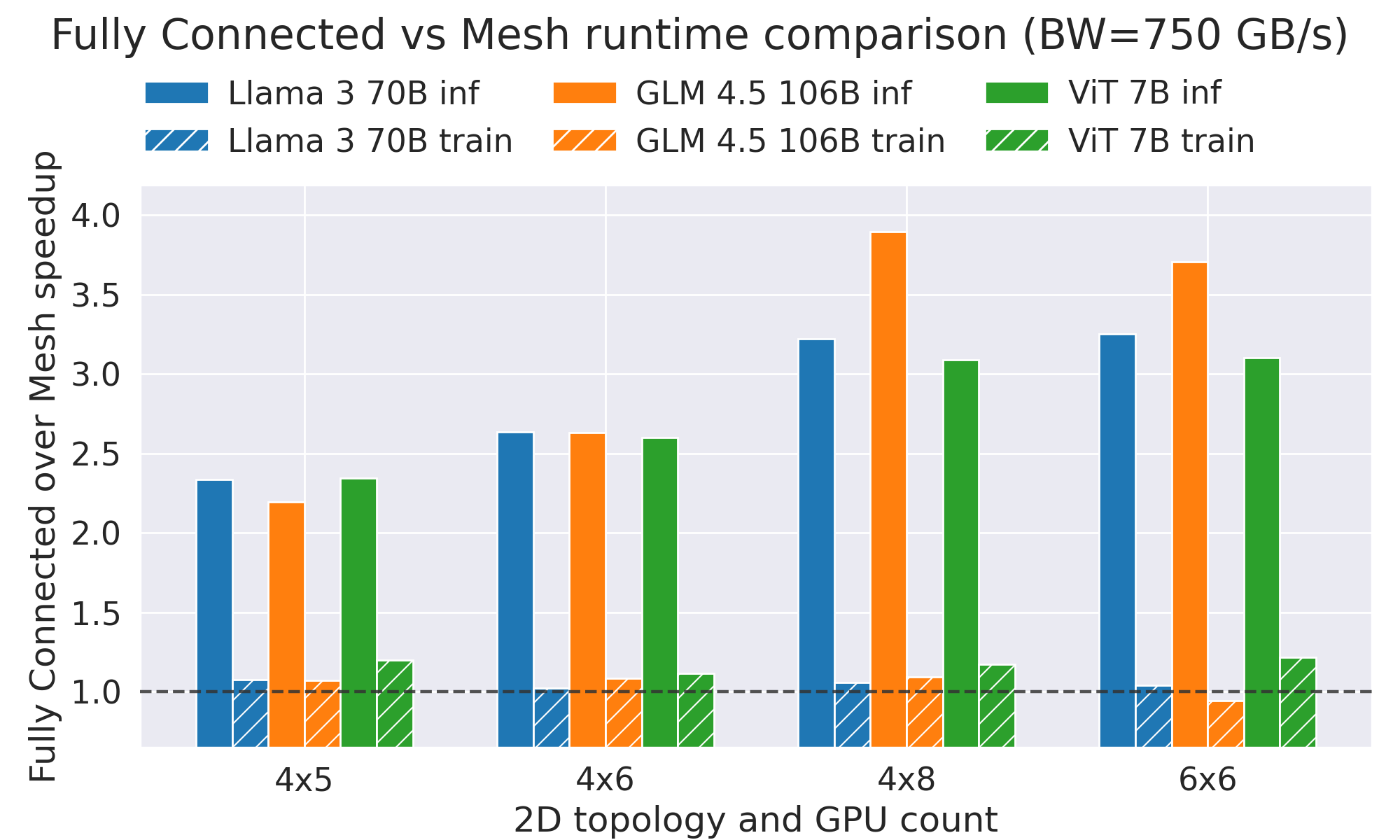}
    \caption{Topology sensitivity on a waferscale-style 2D fabric (750~GB/s per link). We take the baseline 4$\times$5 Mesh of H100-class GPUs from \cite{FRED} and scale to 4$\times$6, 4$\times$8, and 6$\times$6. \textbf{Left:} 2D Torus vs.\ Mesh (wrap-around links). \textbf{Right:} fully connected vs.\ Mesh at matched aggregate per-GPU bandwidth. Bars report speedup relative to the Mesh baseline (higher is better).}
    \Description{Bar charts of Torus-over-Mesh and FullyConnected-over-Mesh speedups for Llama3.1-70B, GLM-4.5-Air, and ViT 7B inference and training at several waferscale system sizes; richer topologies speed up inference substantially while training is largely unchanged.}
    \label{fig:wafer_topology}
\end{figure}
We also study how network topology affects LLM training and inference on a waferscale-style system. The base system is a 4$\times$5 waferscale 2D Mesh of H100-class GPUs with 750~GB/s per link, following \cite{FRED}. We then consider two alternatives: a 2D Torus with wrap-around links at the same link bandwidth, as well as a fully connected fabric approximating aggressive waferscale interconnect proposals enabled by photonic links \cite{ZhangZNSWLQCTPBRE23}. RAPID-LLM models all three topologies explicitly and searches over memory-feasible hybrid-parallel configurations (DP/TP/PP/CP/EP) for each topology and system size, selecting the configuration that minimizes predicted runtime.

Figure~\ref{fig:wafer_topology} reports the Torus-over-Mesh and FullyConnected-over-Mesh speedups for Llama3.1-70B, GLM-4.5-Air \cite{GLMAir} (106B), and ViT 7B \cite{simeoni2025dinov3} under both inference and training. Llama and ViT are dominated by reduction-style collectives (AllReduce/ReduceScatter/AllGather), while GLM also exhibits much heavier expert-routing All-to-All collectives. Both richer topologies help inference and neither helps training, for the same reason: inference collectives carry small messages whose cost is set by hop count and congestion, which added connectivity removes, whereas training's large transfers are bandwidth-bound and gain less given our aggregate bandwidth matching assumption. The torus improves inference for every model, consistently more for Llama and ViT than for the MoE workload, since its wrap-around links shorten paths without substantively increasing congestion-aware bandwidth. The fully connected fabric is far stronger, up to 3.2$\times$ for Llama and 3.9$\times$ for the MoE model. Here, as congestion is significantly reduced, the MoE model is able to benefit as much (or more) than the dense models despite the matched bandwidth assumption.

This study showcases RAPID-LLM's ability to aid in network and hardware design space exploration for different types of inference/training workloads.

\subsection{Hardware what-if analysis for 3D-stacked HBM-on-GPU}
\begin{figure}[!htbp]
    \centering
    \includegraphics[width=\linewidth]{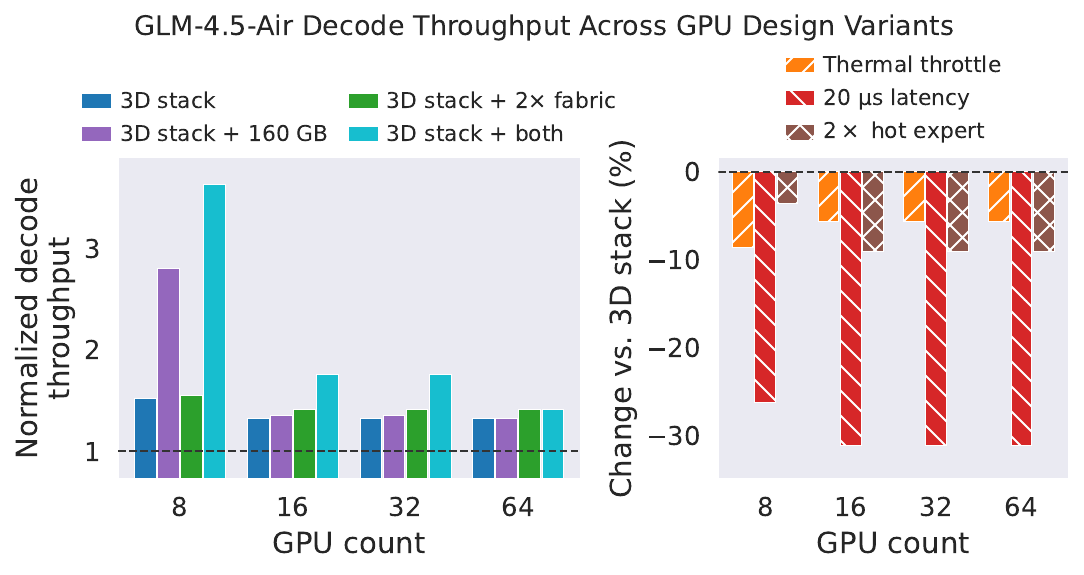}
    \caption{GLM-4.5-Air decode throughput for hypothetical 3D-stacked HBM-on-GPU designs, normalized to the 80~GB 3D baseline at each GPU count. All cases use 3D-stacked HBM with 3$\times$ HBM bandwidth. \textbf{Left:} additional capacity, fabric bandwidth, and their combination. \textbf{Right:} penalties from an HBM thermal throttle, a fourfold increase in scale-up latency to 20~$\mu$s, and a $2\times$ hot expert.}
    \Description{Bar charts of normalized GLM-4.5-Air decode throughput for hypothetical 3D-stacked HBM GPU variants; the left panel shows gains from added memory capacity and fabric bandwidth, and the right panel shows penalties from thermal throttling, higher scale-up latency, and a hot expert.}
    \label{fig:hardware_result}
\end{figure}

For the final case study, we consider a practical system design space exploration problem. 3D Stacking HBM directly on the GPU could provide substantially higher memory bandwidth, but its system-level value depends on the surrounding hardware design. 
We use RAPID-LLM as a hardware ``what-if'' tool to examine a hypothetical H100-like GPU that utilizes 3D packaging to stack its HBM on top of the GPU, giving it three times the baseline HBM bandwidth. We study how memory capacity, fabric provisioning, and plausible implementation constraints affect GLM-4.5-Air \cite{GLMAir} inference. At each GPU count, all results are normalized to the best memory-feasible configuration on the 80~GB 2.5D H100 baseline.

We use a hierarchical ring network with TP/CP/EP on a 100~GB/s scale-up ring and PP/DP on a 25~GB/s scale-out ring.

The left panel explores design choices that complement the 3D stack. Increasing capacity to 160~GB improves throughput by 84.3\% at 8 GPUs because it enables a different parallelism mapping, but adds at most 2.1\% at larger GPU counts. Doubling fabric bandwidth alone adds only up to 6.1\%. Combining both changes produces a larger interaction: throughput improves by 139.3\% at 8 GPUs and 32.6\% at 16--32 GPUs relative to the 3D baseline, because the capacity-enabled mapping exposes communication that benefits from the faster fabric. 

Prospective hardware studies must also test whether projected gains survive physical constraints and workload variability. The right panel therefore applies two hardware nonidealities and one workload stressor to the same 80~GB 3D baseline. 3D stacking HBM-on-GPU can result in thermal throttling that reduces stacked-memory bandwidth down to 73\% of nominal \cite{embisoldram}. This lowers throughput by up to 8.6\%. Increasing scale-up latency fourfold to 20~$\mu$s, approximating a cable-connected rather than tightly integrated domain, lowers throughput by up to 31.2\%. Finally, the balanced-routing baseline is replaced with a skewed case that conserves the total routed-token count but assigns one expert twice the balanced-average load in every MoE layer. This induces imbalanced compute and network congestion, lowering throughput up to 9.1\%.

This study illustrates how RAPID-LLM can evaluate a prospective memory technology together with the capacity, interconnect, physical constraints, and workload variability that determine its system-level value, even for prospective hardware systems that are not available today. 

\section{Limitations and Conclusion}

\subsection{Limitations}
RAPID-LLM is designed for fast design-space exploration, not cycle-accurate prediction. Its frontend uses an analytical extended-roofline operator model. This captures shape- and memory-hierarchy-driven kernel efficiency, but abstracts away compiler-specific behaviors such as exact fusion choices and software serving effects.
For example, NVIDIA's NIM LLM benchmarking report \cite{nvidiaNIM} shows that, for the 5k-input/500-output setting on an H100 80GB GPU, measured throughput under high concurrency changes up to 11\% when software version changes from 1.3.0 to 1.8.0. This difference cannot be captured by analytical operator-level modeling.

On the network side, the backend is analytical and congestion-aware with explicit multidimensional links and fault-aware routing, but it does not model all packet-level dynamics (e.g., transport-level congestion control and detailed switch-buffer behavior). Likewise, our fault model captures soft and hard link faults and their rerouting consequences, but does not represent correlated or time-varying failures such as bursts, multi-link events, or switch failures.
\subsection{Conclusion}
RAPID-LLM is a unified performance modeling framework for LLM training and inference that couples a frontend producing hardware-aware, operator-level Chakra traces with an extended Astra-Sim backend that models explicit multidimensional networks with congestion-aware routing and support for degraded and failed links. Across 124 A100- and H100-based validation cases spanning dense and mixture-of-experts workloads, RAPID-LLM predicts end-to-end inference latency and training time per batch with a mean absolute percentage error of 10.0\%. By enabling fast sweeps over hybrid-parallel configurations and explicit exploration of topology choices, fault scenarios, and GPU/memory what-if designs, RAPID-LLM makes it practical to answer system-design questions that are difficult for trace replay and too expensive for packet-level or cycle-level simulation at cluster scale.
RAPID-LLM is available at \url{https://github.com/nanocad-lab/Rapid-LLM}.
\noindent\paragraph{Future work.}
We see two promising directions: 
\begin{enumerate}
    \item richer inference-serving models, including batching policies and KV-cache management effects~\cite{yu2022orca,kwon2023pagedattention}. 
    \item deeper support for MoE and expert-parallel workloads, including modeling serving strategies such as prefill--decode~\cite{zhong2024distserve} and attention--FFN~\cite{zhu2025megascaleinfer} disaggregation
\end{enumerate}

\appendix
\section{Inference Tool Comparisons}
\label{app:inf_comparison}
GenZ is run at the empirically measured system efficiency its authors report for each hardware configuration~\cite{genz}: 0.4 on A100, and on H100 a value that varies with the number of GPUs (0.55, 0.64, 0.66, and 0.75 for 1, 2, 4, and 8 devices respectively). All other settings are at tool defaults. Vidur is run on its shipped profiling data for valid configurations.

\paragraph{LLMCompass.}
LLMCompass is a well-known analytical hardware-aware prediction model~\cite{zhang2024llmcompass}, making it a natural candidate for comparison. However, it reaches a MAPE of $97.9\%$ on the A100 suite, overestimating throughout, as it does not support Grouped-Query-Attention (GQA) that Llama3 models use. 

\section{Training Tool Comparisons}
\label{app:train_comparison}
STAGE takes activation recomputation as a configuration parameter. After running STAGE with recomputation enabled and disabled, STAGE produced identical ET files and estimated runtime results in both cases. This explains the large MAPE that STAGE reports on these cases.

\paragraph{llm-analysis post-emission correction.}
We run llm-analysis at its own documented settings: the shipped A100-SXM-80GB device specification, its recommended large-scale-training FLOPS efficiency of 0.5, and default memory and communication efficiencies. llm-analysis floors each transformer layer's forward latency at the time to all-gather that layer's weights across data-parallel ranks, $t_{\mathrm{fwd}} = \max(t^{\mathrm{comp}}_{\mathrm{fwd}},\, t^{\mathrm{dp\text{-}ag}}_{\mathrm{fwd}})$, modeling ZeRO-3-style weight unsharding, and applies this floor at any data-parallel degree regardless of the tool's sharding setting, whereas our reference runs use unsharded data parallelism in which no such all-gather occurs. We therefore correct the emitted per-iteration latency using only quantities the tool reports in its own output summary:
\begin{equation}
t_{\mathrm{corr}} \,=\, t_{\mathrm{iter}} \,-\, 3\, n_{\mathrm{acc}} \cdot \max\!\left(0,\, t^{\mathrm{dp\text{-}ag}}_{\mathrm{fwd}} - t^{\mathrm{comp}}_{\mathrm{fwd}}\right),
\label{eq:llma_correction}
\end{equation}
where $t^{\mathrm{comp}}_{\mathrm{fwd}} = t_{\mathrm{attn}} + t_{\mathrm{mlp}} + t_{\mathrm{ln}} + t_{\mathrm{tp}}$ is the reported per-microbatch forward compute and tensor-parallel communication time, $n_{\mathrm{acc}}$ is the number of gradient-accumulation steps, and the factor of 3 reflects the tool's modeling of the backward pass as twice the forward. The correction leaves single-replica ($\mathrm{dp}=1$) configurations unchanged. Without it, llm-analysis reaches 68.2\% MAPE on this suite.

\bibliographystyle{ACM-Reference-Format}
\bibliography{refs}

\end{document}